\documentclass[usenatbib]{mnras} \voffset=-0.5in

\usepackage{graphicx} 
\usepackage{amsmath, siunitx} 
\usepackage{amssymb}
\usepackage{exscale, relsize}
\usepackage{amscd}
\usepackage{natbib}
\usepackage{lscape } 
\usepackage{afterpage}

\usepackage{times}

\usepackage{hyperref}
\hypersetup{colorlinks=true, linkcolor=blue, bookmarks=true, citecolor=black}

\newcommand{\eg}{\textit{e.g.}} 
\newcommand{\ie}{\textit{i.e.}} 
\newcommand{\viz}{\textit{viz.}}

\newcommand{\Ell}{\mathcal{L}}

\renewcommand{\eqref}[1]{\hyperref[eq:#1]{Eq.\ (\ref*{eq:#1})}}
\newcommand{\secref}[1]{\hyperref[sec:#1]{\textsection \ref*{sec:#1}}}
\newcommand{\tabref}[1]{Table\ \ref{tab:#1}}
\newcommand{\figref}[1]{Fig.\ \ref{fig:#1}}

\renewcommand{\sol}{$_\odot$}
\newcommand{\esd}{\Delta\Sigma}
\newcommand{\sigmacrit}{\Sigma_\mathrm{crit}}
\newcommand{\halo}{_\mathrm{halo}}
\newcommand{\nfw}{_\mathrm{NFW}}
\newcommand{\mhalo}{M_\mathrm{halo}}

\newcommand{\Sersic}{S\'ersic}

\newcommand{\sers}{_\mathrm{Sersic}}
\newcommand{\eff}{_e}

\newcommand{\gistar}{(g-i)_*}

\newcommand{\nQ}{\mathtt{nQ}}
\newcommand{\RankBCG}{\mathtt{RankBCG}}
\newcommand{\RankIterCen}{\mathtt{RankIterCen}}

\begin{document}


\title[GAMA+KiDS: Correlations between halo mass and other galaxy
properties]{GAMA+KiDS: Empirical correlations between halo mass and other galaxy properties near the knee of the stellar-to-halo mass relation}


\author[Edward N.\ Taylor et al.]{Edward N.~Taylor,\thanks{entaylor@swin.edu.au}$^{,1}$
 Michelle E.\ Cluver,$^{1,2}$ 
 Alan Duffy,$^{1}$ 
 Pol Gurri,$^{1}$ 
 Henk Hoekstra,$^{3}$\newauthor
 Alessandro Sonnenfeld,$^{3}$ 
 Malcolm N.\ Bremer,$^{4}$ 
 Margot M.\ Brouwer,$^{5}$ \newauthor
 Nora Elisa  Chisari,$^{6}$  
 Andrej Dvornik,$^7$ 
 Thomas Erben,$^8$ 
 Hendrik Hildebrandt,$^7$ \newauthor
 Andrew M.\ Hopkins,$^{9}$ 
 Lee S.\ Kelvin,$^{10}$ 
 Steven Phillipps,$^{4}$ 
 Aaron S.\ G.\ Robotham,$^{11}$ \newauthor
 Cristob\'al Sif\'on,$^{12}$ 
 Mohammadjavad Vakili,$^3$
 and Angus H.\ Wright$^{7}$ \\
$^{1}$Centre for Astrophysics and Supercomputing, Swinburne University of Technology, Hawthorn 3122, Australia\\
$^{2}$Department of Physics and Astronomy, University of the Western Cape, Robert Sobukwe Road, Bellville, South Africa \\
$^3$ Leiden Observatory, Leiden University, PO Box 9513, Leiden, 2300 RA, The Netherlands \\
$^{4}$ HH Wills Physics Laboratory, University of Bristol, Tyndall Avenue, Bristol, BS8 1TL, UK \\
$^5$ Kapteyn Astronomical Institute, University of Groningen, PO Box 800, 9700 AV Groningen, the Netherlands \\
$^6$ Institute for Theoretical Physics, Utrecht University, Princetonplein 5, 3584 CC Utrecht, The Netherlands \\
$^7$ Ruhr-University Bochum, Astronomical Institute, German Centre for Cosmological Lensing, Universitätsstr. 150, 44801 Bochum, Germany \\
$^8$ Argelander-Institut f\"ur Astronomie, Auf dem H\"ugel 71, 53121 Bonn, Germany \\
$^9$ Australian Astronomical Optics, Macquarie University, 105 Delhi Rd, North Ryde, NSW 2113, Australia\\
$^{10}$ Department of Astrophysical Sciences, Princeton University, 4 Ivy Lane, Princeton, NJ 08544, USA \\
$^{11}$ ICRAR, M468, University of Western Australia, Crawley, WA 6009, Australia \\
$^{12}$ Instituto de F\'isica, Pontificia Universidad Cat\'olica de Valpara\'iso, Casilla 4059, Valpara\'iso, Chile \\
}

\maketitle 

\begin{abstract} 
We use KiDS weak lensing data to measure variations in  mean halo mass 
as a function of several key galaxy properties (namely: stellar colour, specific star formation rate, \Sersic\ index, and effective radius)
for a volume-limited sample of GAMA galaxies  in a narrow stellar mass range ($M_* \sim 2$--$5 \times 10^{10}$ M\sol).
This mass range is particularly interesting, inasmuch as it is where bimodalities in galaxy properties are most pronounced, and near to the break in both the galaxy stellar mass function and the stellar-to-halo mass relation (SHMR). 
In this narrow mass range, we find that both size and \Sersic\ index are better predictors of halo mass than either colour or SSFR, with the data showing a slight preference for \Sersic\ index. 
In other words, we find that mean halo mass is more tightly correlated with galaxy structure than either past star formation history or current star formation rate. 
Our results lead to an approximate lower bound on the dispersion in halo masses among $\log M_* \approx {10.5}$ galaxies: we find that the dispersion is $\gtrsim 0.3$ dex. 
This would imply either that offsets from the mean SHMR are closely coupled to size/structure, or that the dispersion in the SHMR is larger than past results have suggested.
Our results thus provide new empirical constraints on the relationship between stellar and halo mass assembly at this particularly interesting mass range.
\end{abstract}


\section{Introduction} 

As the quantitative link between the observed galaxy population and the cosmological population of dark matter halos, the stellar-to-halo mass relation (SHMR) represents a crucial interface between observation and theory \citep[see][for a recent review]{WechslerTinker2018}. 
The messy baryonic processes of galaxy formation and evolution are understood to be seeded by the dissipationless collapse of their larger dark matter halos. 
The ongoing accretion onto and dynamics within galaxies are thus driven by the gravitational potential well at the centre of the halo, and regulated by shocks, outflows, and other gastrophysical processes of feedback within and at the outskirts of the halo. 
Secular evolutionary processes like gas accretion, dynamical instability, star formation, and feedback would lead to the expectation of self-similar evolution of galaxies of a given mass, with the possibility of second order effects tied to formation time, and so to large-scale environment. 
This self-similarity is broken by stochastic, perturbative effects like interactions and/or mergers between galaxies, which can lead to significant differences in the evolutionary trajectories of individual galaxies.
The outstanding challenge of galaxy formation and evolution is to identify and articulate the relative importance of these many different processes and mechanisms by connecting the cosmological population of dark matter halos to the correlated distributions of galaxy parameters as observed in the real Universe. \looseness-1

Techniques like abundance matching \citep[\eg][]{Conroy2006,Guo2010,Moster2010} have been used to derive the average SHMR by forcing consistency between a halo mass function from theory \citep[\eg][]{PressSchechter,ShethTormen1999,Tinker2010} and the observed galaxy stellar mass function \citep[\eg][]{Bell2003,Marchesini2009,Baldry2012,Driver2018}. 
Extensions or refinements like halo occupation modelling \citep[\eg][]{BerlindWeinberg2002,Yang2003} also require consistency with clustering statistics and other observational constraints \citep[but see][who argue that clustering statistics do not have a significant influence on the inferred SHMR]{Moster2010}.
A number of studies have now extended this formalism to infer the SHMR based on the combination of weak lensing measurements with number counts and/or spatial clustering \citep[\eg][]{Leauthaud2012,Velander2014,Coupon2015,vanUitert2016}.
While lensing provides an avenue for direct measurement of the SHMR for $\log M_* \lesssim 10.5$, \citet{vanUitert2016} found that weak lensing data on their own do not provide strong constraints on the SHMR in the high mass regime \citep[but see the recent lensing-only SHMR determination by][]{Dvornik2020}: for $\log M\halo \gtrsim 12$, it is the stellar mass function that provides the tighter constraint on the SHMR. \looseness-1

There is a qualitative and quantitative consensus on the form of the SHMR, at least in terms of a population average, that emerges from these analyses.
The generic result is that the knee in the galaxy stellar mass function is tied to a break in the SHMR around $\log M_* \sim 10.5$, with an associated peak in the stellar-to-halo mass ratio of $\sim $2--3\% and a halo mass $\log M\halo \sim 12$. \citep[\eg][]{Moster2010,Behroozi2010,vanUitert2016}.
On either side of this peak, the SHMR is reasonably described as a power law, with the low-- and high--mass slopes usually taken as reflecting the suppression of star formation by supernova \citep[\eg][]{Larson1974,McKeeOstriker1977,JoungMacLow2006} and by AGN feedback \citep[\eg][]{Croton2006, Bower2006}, respectively, in the low-- and high--mass regimes \citep[see also, \eg][]{Mitchell2016}. 
But it is worth emphasising that this generic result is a virtually inescapable consequence of trying to reconcile the observed Schechter-like galaxy stellar mass function with a close-to-power-law halo mass function \citep[different versions of this argument can be found in, \eg,][]{MarinoniHudson2002, Moster2010, Behroozi2010, vanUitert2016}. Any model that gets both the stellar and halo mass functions right will necessarily give a similar form for the SHMR. \looseness-1

Weak gravitational lensing \citep[see reviews by][]{BartelmannSchneider,HoekstraJain}, and more specifically galaxy--galaxy weak lensing, is one of the most successful observational avenues to obtaining direct halo mass measurements for large and representative galaxy samples \citep[\eg][]{Brainerd1996,Hudson1998,Hoekstra2004,Mandelbaum2006}.
In order to more directly challenge models of galaxy formation and evolution in a cosmological context, our goal in this paper is to take a more empirical approach to exploring the role of halo mass, as measured by galaxy-galaxy weak lensing, in influencing or determining the observable properties of galaxies. \looseness-1

Our specific interest is to probe correlations between halo mass and galaxy properties {\em at fixed stellar mass} --- or, in other words, to identify which galaxy property or properties are most directly correlated with the dispersion around the average SHMR.\footnote{In more theory-minded approaches like abundance matching and halo occupation modelling, the dispersion in the SHMR is usually framed in terms of the distribution of galaxy stellar mass values at fixed halo mass; \ie, $n(M_*|M\halo)$.
For this paper, we consider instead the complementary quantity: the distribution of halo mass at fixed stellar mass, $n(M\halo|M_*$). The two quantities are related, but distinct, with the latter being a more natural observable.} 
By attempting a systematic (if not exhaustive) exploration of second-order correlations around the SHMR, our goals are similar to, but distinct from, studies by \citet{Mandelbaum2006}, \citet{Hudson2015}, \citet{Charlton2017}, and others who have derived SHMRs separately for different galaxy subsamples selected by colour and/or size. 
Our goals are similarly complementary to, \eg, \citet{vanUitert2013}, who have considered whether halo mass correlates more strongly with stellar mass or velocity dispersion \citep[see also, \eg][]{Li2013}.

One novel aspect of this paper is that, as a means to control for the mass dependence of the SHMR, we focus on a narrow mass range in stellar mass: $10.3 < \log M_* < 10.7$, or $M_* \approx$ 2--5 $\times 10^{10}$ M\sol. 
This mass range is particularly interesting for several reasons. 
First, it is close to the knee of both the galaxy stellar mass function and the SHMR.  
It is thus where the stellar-to-halo mass ratio peaks, and so (in the canonical view) the point of transition where stellar feedback gives way to AGN feedback as the dominant regulator of star formation.
\citet{Robotham2014} have also shown, based on galaxy pair counts and star formation rates, that this mass range is where galaxy mergers take over from star formation as the dominant channel for galaxy stellar mass growth.

Perhaps most significantly for this work, this mass range is also where the bimodality (or bimodalties) in galaxy properties is most pronounced, and where there are approximately equal numbers of canonically `early' and `late' type galaxies --- whether that distinction be made on the basis of broadband colour \citep[see, \eg][]{Baldry2006,Peng2012,Taylor2015}, specific star formation rate \citep{RenziniPeng2015}, morphological classification \citep{Bamford2009, Kelvin2014, Moffett2016}, structure \citep{vanderWel2008}, size \citep{Shen2003, Lange2015}, etc.\ \citep[see also, \eg,][]{Robotham2013}.
By having a relatively large spread in galaxy properties across our sample, we obtain the best lever arm on any dependence on halo mass with these properties. 
Further, because this is the mass regime where there is the greatest diversity in galaxy properties, this is also where the influence of halo mass is potentially the most interesting: 
can one or more manifestations of galaxy bimodality be linked to differences in halo mass? \looseness-1

The structure of our discussion is as follows. We lay out our experimental design in \secref{design}, including our sample selection (\secref{sample}) and subdivision (\secref{subsamples}), weak lensing measurements (\secref{lensing}), and halo mass modelling (\secref{fitting}). We present proof of concept for our novel approach in \secref{proof}, including demonstrated consistency with existing results (\secref{consistency}) and a variety of null results (\secref{sanity}), before presenting our main results in \secref{results}. In \secref{scatter}, we discuss our results and their implication for the role of halo mass in galaxy formation and the dispersion in the SHMR. We also consider potential confounding effects and biases in \secref{masserrors} and \secref{concmass}. A full summary of our quantitative results is given in \tabref{summary}, and we summarise our main results and conclusions in \secref{summary}. For the purpose of stellar mass estimates, we assume a \citet{Chabrier2003} stellar initial mass function (IMF), and we have adopted a concordance cosmology ($\Omega_m$, $\Omega_\Lambda$, $h$) = (0.3, 0.7, 0.7) throughout. 

\section{Experimental Design} \label{sec:design}

\subsection{Lens galaxy sample selection} \label{sec:sample}

Our lens sample is selected from the Galaxy And Mass Assembly (GAMA) survey
\citep{Driver2011,Liske2015,Baldry2018}, which has obtained near-total
($\gtrsim 99.5$ \%) spectroscopic redshift completeness for $r < 19.8$ galaxies over
three equatorial fields totalling 180 sq.\ degrees (plus two Southern fields we
do not consider here). We make use of a number of data products that have been
described elsewhere, and have been made public with GAMA Data Releases 2 and 3
\citep{Liske2015,Baldry2018}, including stellar mass estimates and stellar population parameters \citep{Taylor2011}, group identifications
\citep{Robotham2011}, and \Sersic\ profile fits \citep{Kelvin2012}. We also
make use of ultraviolet-plus-total-infrared star formation rates (SFRs)
described in \citet{Davies2016}. \looseness-1

Our primary sample selection is in terms of stellar mass: namely, $10.3 < \log M_*
< 10.7$. The stellar mass estimates are based on stellar population synthesis
modelling of optical--to--near infrared spectral energy distributions (SEDs), using the
\citet{BC03} simple stellar population models with the \citet{Chabrier2003}
prescription for the stellar initial mass function (IMF), and single-screen dust following
\citet{Calzetti}. 
The GAMA SEDs are all measured in (large) matched apertures on
seeing-matched imaging, in order to obtain the best characterisation of SED
shape \citep{Hill2011, Wright2016}. Because the apertures are finite, we
re-normalise the SEDs to match the \Sersic\ model photometry in the $r$-band, as a measure of total flux.
To identify and exclude catastrophic errors in the photometry (which sometimes
happen when a very bright star disrupts the segmentation and aperture
definition) we throw away 185 cases where the \Sersic\ and aperture photometry
are inconsistent, with a difference of more than 0.3 mag. 

\begin{figure*} \centering
\includegraphics[width=17.8cm]{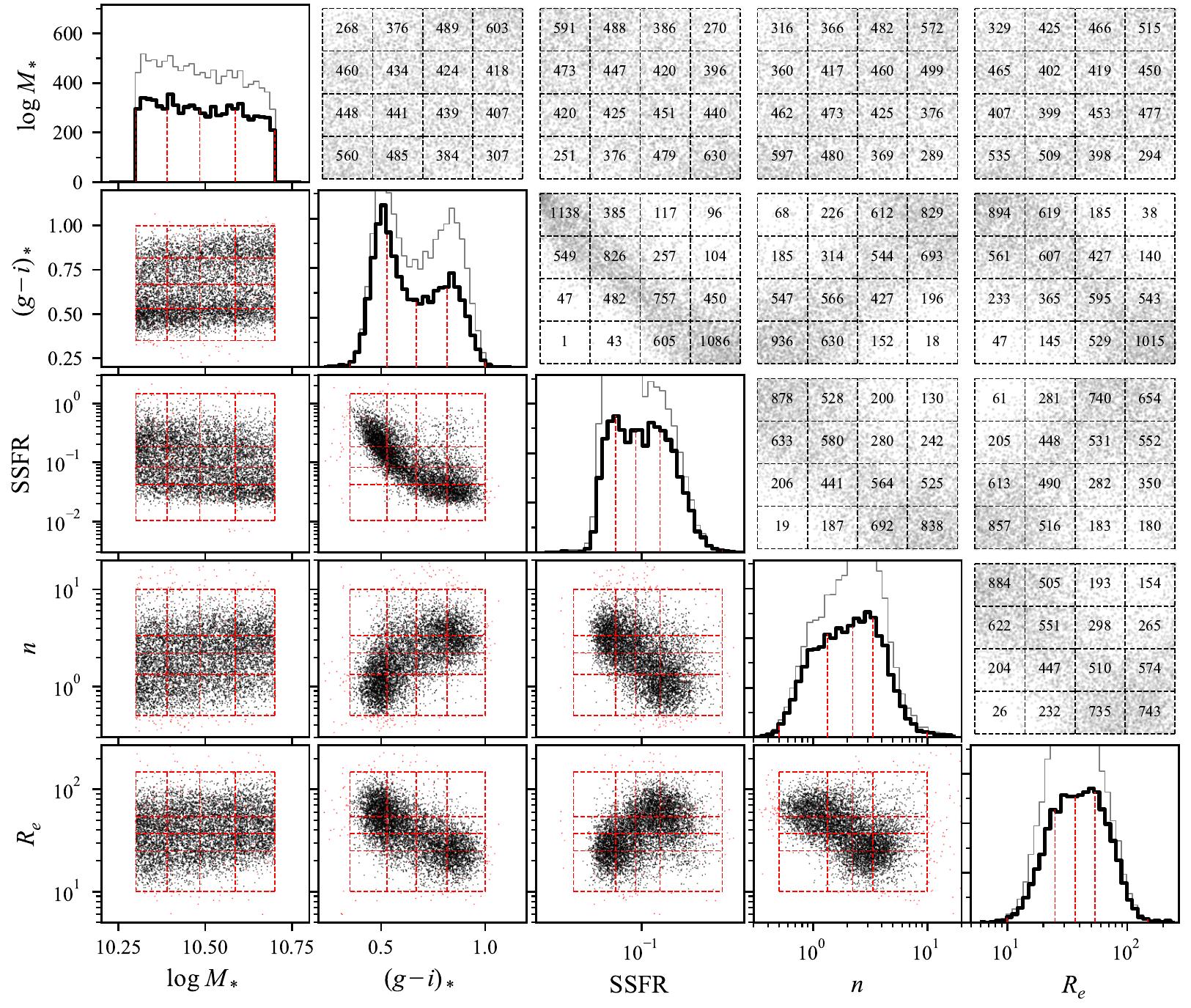}
\caption{Illustrating the distributions of and correlations between galaxy
parameters within our sample of $\log M_* \sim 10.5$ galaxies.--- Along the
diagonal, panels show histograms of the distributions of: stellar mass, $\log
M_*$; intrinsic (dust-corrected) stellar colour, $\gistar$; specific star
formation rate, SSFR; \Sersic\ index, $n$; and half-light radius, $R_e$. The
black histograms show our sample of central galaxies only; the lighter gray
histogram shows the parent sample, including satellites. In the lower left
panels, the black points show bivariate distributions for these parameters. The
red lines show how the sample is divided into four equally-sized samples
according to each property, after excluding a few outlying values (red points).
The upper right panels explicitly show the overlap between these different
sample divisions. As discussed in \secref{subsamples}, the main points to take
from this figure are 1.)\ across this limited mass range, the distribution of
other galaxy properties as a function of mass is roughly constant (each cell in
the top row is roughly uniformly populated); and 2.)\ although there is strong
covariance between other galaxy properties (there is clear structure in the
gray points), there are considerable differences when dividing the sample
according to different properties (many galaxies are found in off-diagonal
cells). \label{fig:sample}} \end{figure*}

We limit our sample to the redshift range $0.10 < z < 0.18$. The upper redshift
limit, which is to ensure that we have a properly volume-limited sample, has been
determined following the arguments given in \citet{Taylor2015}; see also Fig.\
14 of \citet{Baldry2018}. 
The lower limit is to protect against potential systematics tied to redshift errors, since the usual linear propagation of error/uncertainty can break down at the lowest redshifts. 
With the selections above and the usual quality cut $\nQ \ge 3$ for redshift reliability, this gives a parent sample of 11392 galaxies.

In order to ensure that we are truly looking at the primary halos of the
particular galaxies in our lens sample, we have also done our best to select only
central galaxies based on the GAMA group catalogues \citep[G$^3$C;][]{Robotham2011}.
First, if a galaxy is ungrouped in the G$^3$C then it is taken to be a
central by default. Then, the G$^3$C gives two quantities that
can be used to identify the central galaxy within a group: $\RankBCG$, which
ranks galaxies within each group according to brightness, and $\RankIterCen$,
which ranks galaxies according to their distance from an iteratively
re-calculated centre of light. For our analysis, we require either
$\RankBCG = 1$ and $\RankIterCen < 4$ or vice versa; that is, we require
approximate consistency between the two measures. This gives us our main sample
of 7593 central galaxies. Note that both values are precisely 1 for $\approx
93$ \% of our main sample, and that our final sample is $\approx 3$ \% larger
than it would be if we defined our sample based on just one of the two
measures. In any case, we have repeated our analysis using different central
galaxy selections and verified that none of our main results or conclusions
change.

\subsection{Sample subdivision} \label{sec:subsamples}

\figref{sample} shows the range of properties spanned by the galaxies in our
sample --- namely: stellar mass, $\log M_*$; intrinsic (\ie, corrected for internal dust attenuation)
stellar colour, $\gistar$; specific star formation rate (SSFR); \Sersic\ index,
$n$; and \Sersic\ effective radius, $R_e$. In the diagonal panels of this
Figure, the lighter gray histogram shows the distributions for the parent
sample, and the black histograms refer to our main sample of central galaxies
only. We note that while the effect of excluding satellite galaxies in this
mass range is to slightly reduce the relative numbers of generically red,
passive, and de Vaucouleurs-like ($n \gtrsim 2$) galaxies, it also leads to a
more nearly flat distribution of $\log M_*$ for our main sample.

One of the ways that we will explore halo mass variations will be to split our
main sample into subsamples, according to a particular property. The dashed
lines in each panel of \figref{sample} show the quartiles for each property
within our main sample; \ie, these lines show how to split our main sample into
four equally-sized subsamples on the basis of any one property. One nice aspect
of selecting this particular mass range is that dividing this sample into
quartiles closely aligns with the peaks and saddle of the bimodalities in
colour, SSFR, shape, size, etc.

The principal difficulty that we will grapple with in this paper is that, even
at fixed mass, many of these properties are closely correlated. There are good
astrophysical reasons why a sample of galaxies with blue colours is also
likely, in general, to have higher star formation rates, and to have a diskier
morphologies (and hence lower values for the \Sersic\ index, $n$). This can be
seen from the bivariate distributions for the main parameters of interest
within our main sample, which are shown in the off-diagonal panels of
\figref{sample}. As a quantitative description of the overlap between
subsamples divided in different ways, the upper-right panels of \figref{sample}
remap each property to a dimensionless rank or percentile. Each cell in these
panels thus shows how the quartile subdivision in one property projects onto a
similar subdivision in the other property; the numbers refer to how many
galaxies are found in each cell. For example, it can be seen that while the
bluest quarter of our main sample (the lefthand column in the upper
$\gistar$--$\log M_*$ panel) do span the full mass range that we consider, it
is nevertheless biased slightly to lower stellar masses (more galaxies in the
lower cells than in the upper cells); conversely, the reddest quarter of the
sample (righthand column) has slightly higher stellar masses (more galaxies in
higher cells). Similarly, the lowest quarter of our sample in stellar mass has
on average bluer stellar colours, and the highest quarter in stellar mass has
on average slightly redder stellar colours. 

By virtue of our decision to focus on only a narrow stellar mass range,  the interdependence is stronger between parameters other than mass. 
This shows how we can effectively control for the mass dependence of the SHMR to isolate second order correlations between halo mass and the dispersion in the SHMR.
For the other properties we consider, the interdependencies mean that any `true', causal
relation (or relations) between halo mass and one (or more) of these properties
will induce `spurious', coincident correlations with other properties. 
On the
other hand: while these interdependencies are significant, they are not total
-- there are real differences between subdividing this sample according to
different properties. This gives us cause to hope that we may be able to
distinguish between a genuine correlation between lensing signal and some
galaxy property and any `spurious' or tertiary correlations. 
We return to this point in
\secref{scatter}.

\subsection{Weak lensing measurements from KiDS} \label{sec:lensing}

The weak lensing measurements that are the focus of this paper are derived from the shapes of much more numerous background source galaxies, as measured in $ugri$ optical imaging from the Kilo Degree Suvey \citep[KiDS][]{Kuijken2015,deJong2017,Hildebrandt2017, Hildebrandt2020}.  
The KiDS source catalogues are composed of $\sim$15 million $z \lesssim 1.2$ galaxies, almost fully encompassing the (much smaller) GAMA fields.
The distorting effect of weak gravitational lensing, called shear, is to slightly change the observed ellipticity and position angle of a background source seen in close projection to a nearer lens. 
The aggregate effect for an ensemble of many background sources is that their sizes are seen to be, on average, very slightly compressed radially and stretched tangentially around the lens.

In the thin lens approximation, and assuming circular symmetry for the projected lensing
mass, the degree of shear is related to the geometry of the
observer--lens--source configuration, and to the mass distribution of the lens
via:
\begin{equation} \label{eq:gamma}
\gamma(R) = \esd(R) \, / \, \sigmacrit ~ .
\end{equation}
The action of the lensing mass distribution is determined via the {\em excess surface density}
(ESD), which can be expressed as:
\begin{equation}
\esd(R) = {\widehat{ \Sigma }(R) - \Sigma(R)} ~ .
\end{equation}
In words, this is the difference between the mean projected surface density
within the radius $R$:
\begin{equation}
   \widehat{ \Sigma }(R) = {1 \over \pi R^2} \int_0^R \, dR' \, \Sigma(R')
\end{equation}
and the projected surface density at that radius, $\Sigma(R)$. All else
being equal, the degree of shear thus depends on the {\em density contrast},
and not the density (or mass) {\em per se}.
The gravitational and geometric dependence of the shear is fully encapsulated by 
the critical surface density, $\sigmacrit$, which is given by:
\begin{equation} \label{eq:sigmacrit}
\sigmacrit = {c^2 \over 4 \pi \mathrm{G}} ~ { D_l \over D_s \, D_{ls} } ~ ,
\end{equation}
where $D_l$, $D_s$, and $D_{ls}$ are the angular diameter distances between the
observer and the lens, between the observer and the source, and between the
lens and the source, respectively. 

The method for deriving weak lensing measurements from KiDS imaging is
described in detail in, e.g.,\ \citet{Viola2015}, \citet{Hildebrandt2017}, and \citet{Dvornik2018}.
In brief: 
galaxy shape measurements are made using the \emph{lens}fit method \citep{Miller2013}, and then the lensing signal is
measured in annuli as a weighted average of the tangential projections of the background
source ellipticities, to build up a lensing profile for each lens; \viz:
\begin{equation} \label{eq:esd}
\esd_{R,i} = \left({
 \sum_{j} ~ \widetilde{w}_{ij} ~ \varepsilon_j ~ \widetilde{\Sigma}_{\mathrm{crit},ij} 
 \over \sum_{j} ~ \widetilde{w}_{ij} } \right) 
 ~ {1 \over 1 + K(R)} ~ .
\end{equation}
Here, $\esd_{R,i}$ is the ESD profile for the $i$th lens measured in an annulus
with radius, $R$; $\varepsilon_j$ is the tangential projection of the shape tensor for
the $j$th source; $\widetilde{w}_{ij}$ is the weight given to each source according to
its ellipticity and the lens--source geometry; 
$\widetilde{\Sigma}_{\mathrm{crit},ij}$ is the effective critical surface density
for the lens/source pair $ij$; and $K$ is a small ($\lesssim 10$ \%) scalar
correction to account for the multiplicative `noise bias' and `weight bias' 
in the overall shear inferred from the optimally weighted shapes of
small and/or low signal-to-noise galaxies. 
The \citet{FenechConti2017} calibration of the $K$s used here has recently been updated by \citet{Kannawadi2019}, but the differences are negligible for our purposes.

We use the KiDS galaxy--galaxy weak lensing pipeline \citep[see, \eg][]{Dvornik2018} to obtain ESD
profiles for each of the galaxies in our lens sample, based on the KiDS-450
catalogues \citep{Hildebrandt2017}. This pipeline builds ESD profiles, given lens positions and redshifts, and catalogues of shape and redshift information for background sources, as described in \citep{Kuijken2015}.
Specifically, we measure ESD profiles in 20 concentric annuli around each lens, with a logarithmic spacing between annulus edges in the range 12--2000 kpc (corresponding to $\approx$ \ang{;;5}--\ang{;13;} at the median redshift of our sample), and only
considering background sources with $z_s > 0.2$ and $(z_s - z_l) > 0.1$. 

The source redshift distribution (which enters via the $\widetilde{\Sigma}_{\mathrm{crit},ij}$ in \eqref{esd}) is estimated separately for each foreground lens, using the direct re-weighting approach described in \citet{Hildebrandt2017}, which is itself based on the method proposed by \citet{Lima2008}. 
Similar to the shape measurements, any residual errors/uncertainties in the photometric redshift estimates, which are typically at the level of a few percent, are not formally propagated into the ESD measurements.
We note, however, that to the extent that any such errors are tied to the {\em sources}, and not the {\em lenses}, they will effect all of our galaxy subsamples in the same way.  
Such systematics thus have no impact on our ability to identify relative trends across the sample, which is our principal focus.

In principle, because the same background source galaxy can contribute to the
ESD profile for multiple lenses, the errors in the ESD profiles of different
lenses can be significantly correlated, particularly where there is a high density
of lenses on the sky. In practice, for our specific and relatively small sample
of $\log M_* \sim 10.5$ galaxies, this covariance is negligible: the Pearson
correlation coefficients for ESD profile measurements at different radii for
different lenses are $\lesssim 0.001$. This should not be surprising given the
relatively low sky densities of our lens sample: $\sim 40$ / deg$^2$, over
three independent fields. We therefore ignore the covariances between the ESD
measurements for the lenses in our sample, which makes the computation
described in the next section tractable.

These ESD profiles --- one for each of the lens galaxies we consider --- 
represent the weak lensing dataset that we analyse in this paper.
While the signal-to-noise ratio in any one ESD profile is very low, by considering them aggregate, we can hope to derive information about the global lensing properties of the ensemble.

\begin{figure*} \centering
\includegraphics[width=17.8cm]{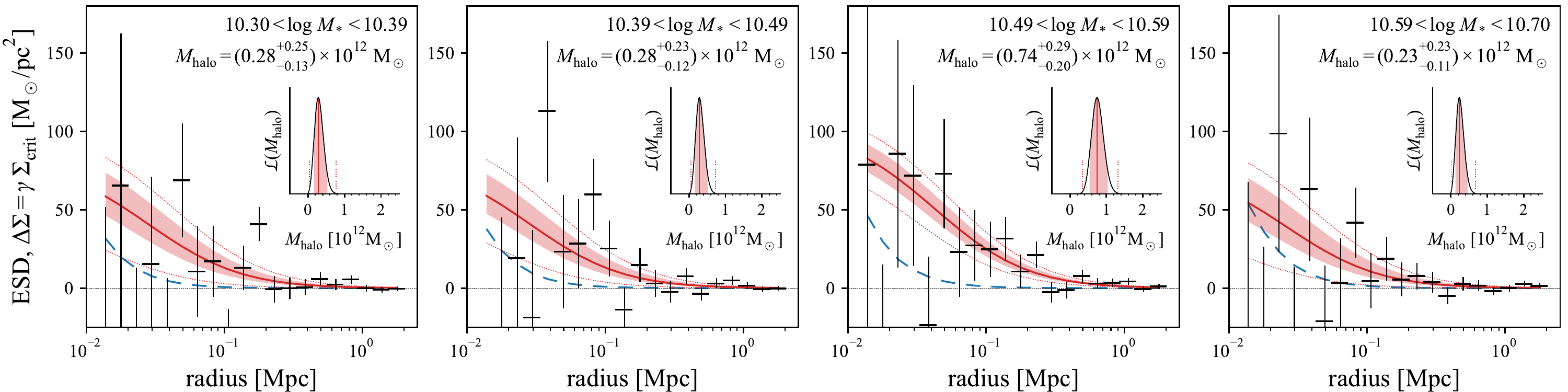} \caption{Stacked
lensing profiles and halo mass modelling, having subdivided our sample
according to stellar mass, $\log M_*$. --- In each panel, the mean lensing
signal from the halo is shown in black; in each case, these are based on the
KiDS lensing measurements for approximately 1600 lenses. The blue dashed line
shows the lensing contribution from the stars, based on the GAMA
\Sersic\ fit parameters for the lens galaxies. These values have been
subtracted from the observed ESD profiles to isolate the effect of the halo. The red
shows the inferred NFW halo model: the heavy solid lines show the maximum
likelihood fits; the shaded regions show the equivalent of the $\pm 1 \sigma$
uncertainties; the thin dotted lines bound the 95 \% confidence region. Inset
within each panel is the posterior PDF for the mean halo mass for each bin.
Note that for each subsample, the uncertainty on the halo mass estimate is
$\approx 0.2 \times 10^{12}$ M\sol. \label{fig:esd_mstar}} \end{figure*}

\subsection{Lens modelling and mass estimation} \label{sec:fitting}

\eqref{gamma} shows how, knowing both the geometry of the lens--source system
and the projected mass distribution of the lens, one can predict the shear.
Conversely, given observational constraints on both the geometry and the shear, one can infer the lens mass
distribution.

We assume that the halos can be described using the NFW mass distribution
\citep{NFW}, which is fully described by its mass\footnote{Following the
convention for $N$-body dark matter simulations, $M_{200}$ is defined as the
mass enclosed within a radius $R_{200}$ from the halo centre, such that the
mean mass density within $R_{200}$ is 200 times the cosmological mean matter density at
that redshift, \ie, $\bar{\rho} = 200 ~ \Omega_m(z) \, \rho_\mathrm{crit}$. With this definition, the shape parameter $c$ can be related to a
scale radius $R_s$ via $c = R_{200} / R_s$. Note that for clarity,
elsewhere in the paper we will use the symbol $\mhalo$ and just `the halo
mass'; this should properly be understood as referring to this proxy
measurement of the total or virialised halo mass.}, $M\halo$, and a shape
parameter, $c$, which is usually referred to as the concentration. 
Analytic expressions for the values of
$\widehat{\Sigma}$, $\Sigma$, and/or $\esd$ for the NFW profile
are given in, \eg, Eq.s 11--15 of \citet{WrightBrainerd} or Eq.s 40--43 of \citet{Coe}.
We find that, in general, we cannot place strong
constraints on the values of $c$ using our data (see \secref{concmass} where we explore this issue in
greater detail). For this reason, and following \citet{vanUitert2016}, we adopt a fiducial
mass-concentration relation based on \citet{Duffy2008}:
\begin{equation}
    c = f_ \mathrm{conc} \, \cdot \, 10.14 \, \left({ M\halo \over 2 \cdot 10^{12}\, h^{-1} \, \mathrm{M}_\odot }\right)^{-0.081} \, (1+z)^{-1.01} ,
\end{equation}
where the scaling factor $f_\mathrm{conc} = 0.70$ comes from the lensing-plus-stellar mass function modelling of \citet{vanUitert2016}. \looseness-1

For small values of $R$, the stars within the galaxy itself can make a
non-negligible contribution to the observed shear. In the thin lens
approximation, the shear from the stars is simply added linearly to that from
the halo. Knowing each galaxy's total stellar mass, circularised effective
radius, and \Sersic\ shape parameter, we can approximately account for this
using a circularised \Sersic\ profile to describe the stellar mass distribution
of the galaxy. The relevant assumptions are 1.)\ that the \Sersic\ parameters
derived from fits to the $r$-band images can be used to describe the stellar
mass distribution, and 2.)\ that a circularised model is sufficient for the
purposes of computing the lensing shear. Analytic expressions for the values of
$\widehat{\Sigma}$  and $\Sigma$ for the circularly-symmetric \Sersic\
profile can be obtained from, \eg, Eq.s 1 and 2 of \citet{GrahamDriver}.

Putting these two pieces together --- \ie, knowing the \Sersic\ parameters for the
galaxy, and given a trial value for the halo mass --- we can then generate a
model ESD profile to compare to the data. The goodness of fit for the ensemble is given by the summation over all radii and for all lenses:
\begin{equation} \begin{split} \label{eq:chi}
    \chi^2(M\halo) = \sum_{R,i} 
    \bigg[ \esd_{R,i} - \esd_{\mathrm{\sers}}(R\,|\,M_{*,i}, R_{e,i}, n_i) \\
 -  \esd\nfw(R \, | \, M\halo) \bigg]^2 / { \sigma_{R,i}^2}    ~ ,
\end{split} \end{equation}
where $\sigma_{R,i}$ is the formal uncertainty associated with the measured
$\esd_{R,i}$. 
Noting that the summation over $i$ can be evaluated for the sample as a whole or for any specific subset of the sample, the (log) likelihood function for the ensemble is then just:
\begin{equation} \label{eq:ell}
    \ln \Ell(M\halo) = {-1 \over 2}  \chi^2(M\halo) ~ .
\end{equation}

Within the framework of Bayesian statistics, the quantity of interest is not
the likelihood {\em per se}, but instead the posterior probability distribution function
(PDF) for the value of $M\halo$, which is given by the product of the
likelihood and an assumed prior on $M\halo$. Here, we adopt a prior that is
flat in $M\halo$ (and not, say, flat in $\log M\halo$) which means that
the PDF is directly proportional to $\Ell$ as defined above. The effect of
choosing a prior that is flat in $\log M\halo$ would be to make the PDF
proportional to $\Ell/M\halo$. With the data used here, this $1/M\halo$
up-weighting blows up more rapidly than the value of $\Ell$ drops for values of
$M\halo \ll 10^{11}$ M\sol . The result is a pathological effect where 
the PDF diverges for small values of $M\halo$. Quite aside from this point,
this choice of prior is also natural since, all else being equal, our
observables (\ie, the ESD and the shear) scale linearly with mass, and our fits
are framed in terms of linear ESD. In other words, our decision to use a flat
prior in $M\halo$ is crucial to our results, and also reasonable.

Looking at \eqref{chi} and \eqref{ell}, it can be seen that if all lens galaxies
are assumed to have the identically the same halo mass, then the summations over $i$ need only
be done once. If this summation is done in advance then the computation of the
likelihood function  involves just the one comparison between that co-added ESD
profile, representing the ensemble in aggregate, and one for the model. 
This is the standard approach of `stacking', which has the well known limitation
that it is necessarily limited to considering the mean properties of the larger
ensemble (and always with the implicit assumption of Gaussian statistics). \looseness-1

To overcome this limitation, we  also pursue a different approach: fitting each lens in the full ensemble simultaneously, while allowing each lens to have its own particular mass. 
To do this, we adopt a simple (linear) parametric prescription to predict halo mass from some other property (\eg,
colour, size, etc.), here denoted $x$: \looseness-1
\begin{equation} \label{eq:mhi}
	M_{\mathrm{halo},i}(x_i | A, \, b) = A \, ( x_i - x_0 ) + b ~ ,
\end{equation}
Being free to choose any value for the arbitrary reference value, $x_0$, we choose $x_0$ to be the mean value of $x$ for the ensemble, so as to minimise the covariance between the parameters $A$ and $b$. 
We neglect any observational errors/uncertainties in the $x$ values, essentially because properly accounting for these errors is computationally expensive. 
The effect of this decision will be to (weakly) systematically bias our results towards apparently weaker correlations; \ie, lower values of $A$. \looseness-1

The only other additional complication is whether and how to accommodate negative values for $M_{\mathrm{halo},i}$.  
Even though, physically, mass is a strictly positive quantity, from an
experimental standpoint it is perfectly reasonable to obtain a negative
measurement where the errors are comparable to the actual value. 
Our simple scheme for treating negative values of $M\halo$ is so use the absolute value of $M\halo$ to determine the shape of the ESD
profile, and then reverse the sign of the shear in the case of a negative mass,
so that $\esd(-M\halo) \equiv - \esd(M\halo)$. The motivation and rationale for this
decision---essentially, to limit potential biases in the inferred values for $A$---are discussed further in \secref{sanity}.

\begin{figure} \centering
\includegraphics[width=8.6cm]{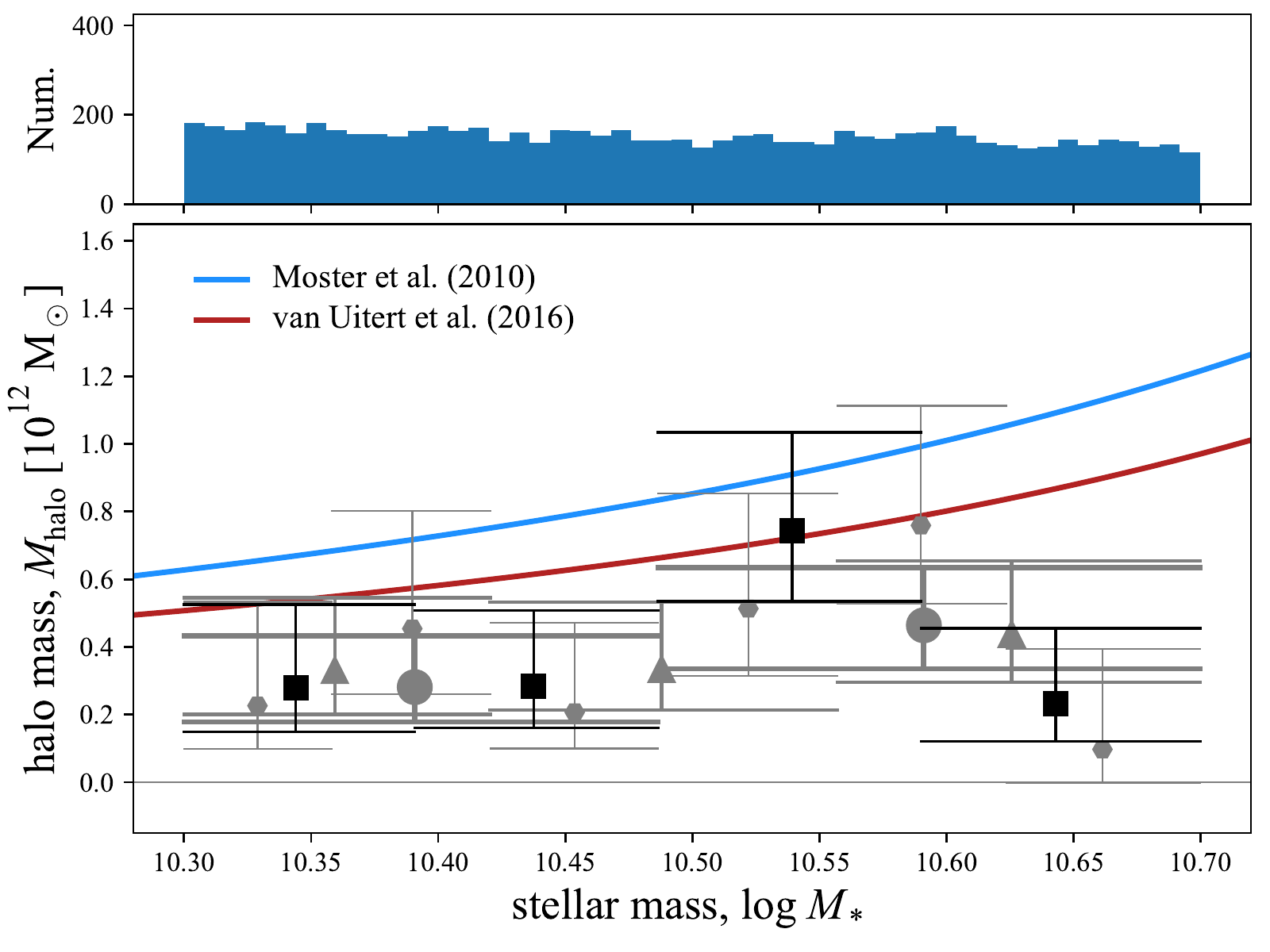}
\caption{Inferred halo masses from stacked ESD profile fitting for our sample
sub-divided by stellar mass.--- The black squares show the inferred halo masses
for the four equally populated bins shown in \figref{esd_mstar}; the gray
points show what we get when splitting our sample into two (circles), three (triangles), or six (hexagons)
equally sized bins. The horizontal length of the errorbar caps show the extent of each
bin, and the size of each point reflects the number of lenses that have gone
into each stack. The points are the maximum likelihood values, and the (asymmetric) error bars reflect the
68 \% confidence interval in the PDF for $M\halo$ for each
stack. 
\label{fig:logmstacks}} \end{figure}

Rather than fitting for the value of a single parameter, $M\halo$, for the
ensemble, we can thus allow each lens to have its own unique halo mass,
$M_{\mathrm{halo},i}$, which is derived from the value of $x$ for that lens,
$x_i$, and the free parameters $A$ and $b$. We adopt the standard
`non-informative' or reference prior, which is flat in both $A$ and $b$. Our
likelihood function then becomes:
\begin{equation}
	    \ln \Ell(A,b) = {-1 \over 2} \, \chi^2(A,b) ~ ,
\end{equation}
with the only other formal modification being to replace the $\mhalo$ that
appears in \eqref{chi} with $M_{\mathrm{halo},i}(x_i |A,b)$ as defined in
\eqref{mhi}. 
Formally, the approach outlined above is equivalent to fitting for the ensemble of values $M_{\mathrm{halo},i}$ for all lenses in the sample (which are in general poorly constrained on their own), and then fitting a linear relation to those results.
Seen through this lens, our approach is not to track the constraints on the individual $M_{\mathrm{halo},i}$ values, and instead marginalise over these unknowns as nuisance parameters.

The practical consequence of this change is that we we cannot pre-compute the
summation over $i$ in the definition of the $\chi^2$: it is necessary to
compute a separate model--observed ESD comparison for each individual lens, and for every set
of $(A,b)$ trial values. 
But the benefit justifies the cost: by allowing each lens to have its own halo mass, it becomes possible to make inferences about the distribution of halo masses across the ensemble --- which, after all, is our primary goal in this paper.

\section{Proof of concept and some simple sanity checks} \label{sec:proof}

\subsection{The stellar-to-halo mass relation across our sample} \label{sec:consistency}

As a demonstration of our ability to identify and measure variations in halo
mass across our sample, we first consider the correlation between stellar mass
and halo mass. A secondary motivation here is to check  that the results
we obtain are broadly consistent with existing results. From the outset,
however, we make the disclaimer that past determinations have been derived 
following very different methods and assumptions, and our ability to constrain the relation
between stellar and halo mass with our sample is limited by our deliberate
decision to focus on a narrow mass range.

\begin{figure} \centering
\includegraphics[width=8.6cm]{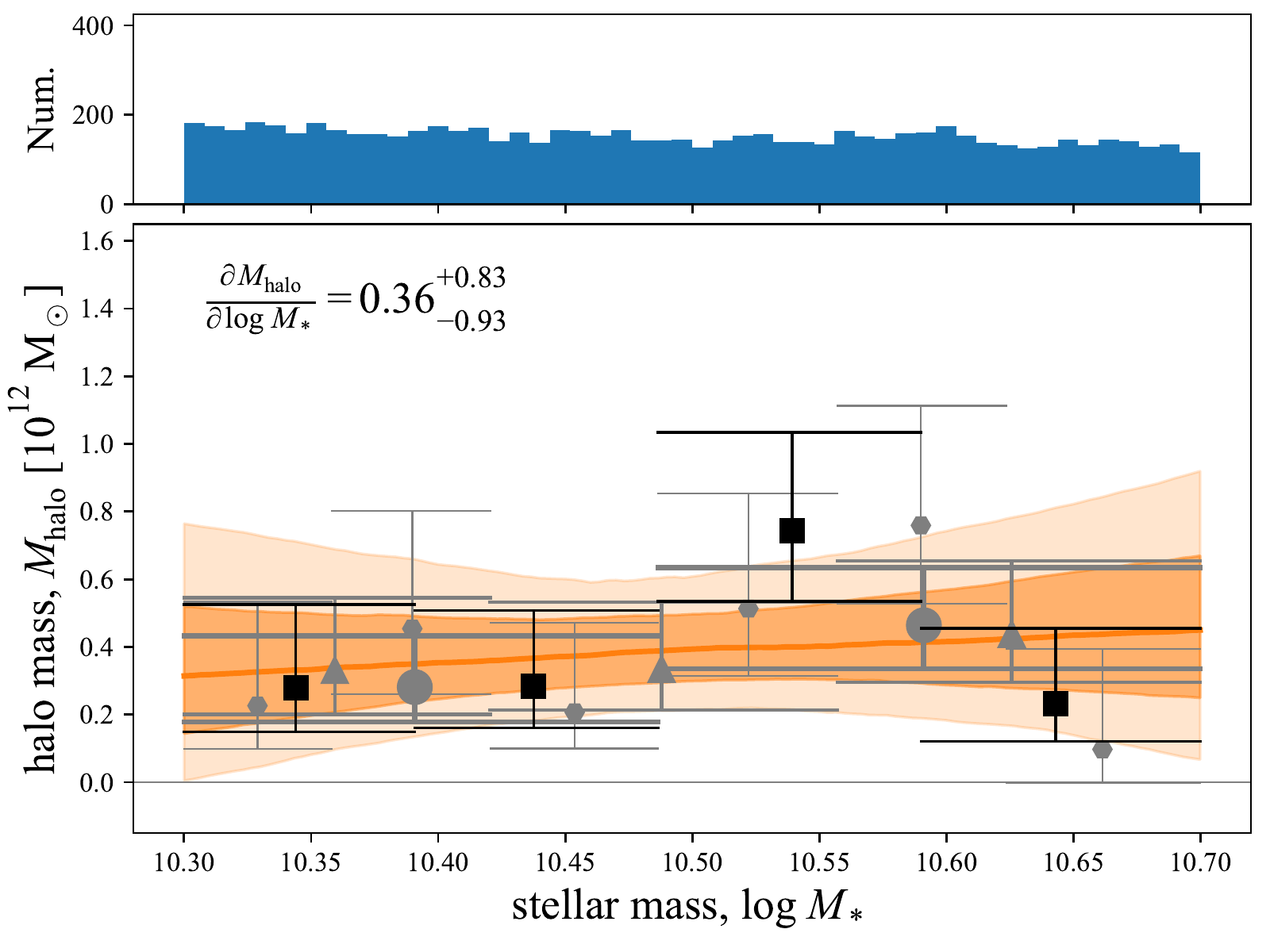}
\caption{Inferred linear correlation between halo mass and stellar mass across our sample.---
The solid yellow line shows the best fit linear relation between $\log M_*$ and
$M\halo$, based on simultaneous fits to the KiDS ESD profiles for our sample of
lenses. The darker and lighter shaded regions show the percentile equivalents
of the $\pm 1$ and $\pm 2 \sigma$ regions for the fit. The points and error
bars are simply repeated from \figref{logmstacks}; the line shown is {\em not}
simply a fit to these points. Instead, the line and the various points are both
different visualisations of the same underlying lensing behaviour of the
galaxies in our sample.\label{fig:afologm}}
\end{figure}

\subsubsection{Stacked lensing profiles} \label{sec:profiles}

\figref{esd_mstar} shows the mean halo lensing profiles for our lens sample,
subdivided by stellar mass into four equally populated bins. 
In each panel of \figref{esd_mstar}, the blue dashed line shows the mean lensing profile for the stellar component, which is derived from the combination of the SED-fit mass-to-light ratios and \Sersic\ fit parameters from GAMA. 
These values have been subtracted from the KiDS lensing measurements, shown in black, to isolate the lensing effect of the halo. 
We note that accounting for lensing by the stars reduces the inferred lensing mass by $\sim$10--15\%; \ie, by $\sim$3--5 times the actual stellar mass.\footnote{The reason for this discrepancy comes down to the different ESD profile shapes for the stars and the halo, as can be seen in \figref{esd_mstar}.  Because the stellar mass profile is steeper than for the dark halo, less stellar mass is required to the produce a similar degree of shear at small radii.  Conversely, at larger radii, the shear signal is completely dominated by the larger and more diffuse halo.} 
Allowing that our simple accounting for the lensing effect of the stars (using single and circularly symmetric \Sersic\ profiles) may only be accurate to within, say, 10--20\%, the impact of these errors on the inferred halo mass will be small: a  10--20\% residual from a 10--15\% effect amounts to only a few percent in the final result.
In other words, while it is important to make some accounting of the lensing effect from the stars, having done so, the residual errors in our results arising from imperfect modelling of the stellar profiles will be negligible in comparison to the statistical uncertainties.

The maximum likelihood estimates for each halo ESD
profile are shown in \figref{esd_mstar} as the red lines, with the red shaded regions showing the equivalent of the $\pm 1
\sigma$ uncertainties (\ie, the 68\% confidence interval), and the red dotted lines bounding the 95\% confidence
region.
The inset panels show the posterior PDF for the mean halo mass for the sample,
with the same confidence limits marked in a similar fashion. Note that simple
Gaussian statistics are not always a good way to represent our results,
particularly where there is small but non-zero likelihood that $M\halo \le 0$.
(Recall that we do allow the parameter $M\halo$ to be negative; see
\secref{fitting} and \secref{sanity}, below.) For example, in the first panel, the PDF for the lensing
halo mass can be seen to be slightly skewed towards higher masses. For this
reason, we will always show the (generally asymmetric) 68\% confidence limits instead of $\pm 1 \sigma$ errors.

\subsubsection{Measuring mean halo mass in bins of fixed stellar mass}

\figref{logmstacks} shows the halo masses we derive from our stacked ESD
profile fits for our sample sub-divided according to stellar mass. The black
points highlight the four equally-sized mass bins shown in \figref{esd_mstar}.
But we could just as well have split our sample into fewer or more bins: the
gray points show what we get splitting our sample into two, three, or six
equally populated bins. 
(For reference, the histogram in the upper panel shows the observed $\log M_*$ distribution for our sample.)
Naturally, the uncertainties are larger for the smaller
and less populated bins. Since each set of points is depicting the same data,
it is not surprising that they all show a more or less consistent picture. That
is, these binned results are a useful means for visualising the average
relation between halo mass and stellar mass across our sample.

For comparison, we show the SHMR determination from \citet{vanUitert2016}, which is derived from halo modelling of the joint GAMA+KiDS dataset over 100 square degrees, and also the one by \citet{Moster2010}, which is derived from halo occupation modelling on SDSS data.  
Noting that the formal statistical uncertainties on each of these curves is at the level of $\sim 40$\%, our directly inferred halo mass measurements are slightly but systematically low compared to the two SHMRs shown here: where we might have expected a mean halo mass of $\sim (1 \pm 0.1) \times 10^{12}$ M\sol, the mean measured value for our sample is $(0.4 \pm 0.1) \times 10^{12}$ M\sol. 
It is beyond the scope of this paper for us to resolve this apparent disagreement, but we note that where our results come from direct observational constraints on the gravitational mass surrounding our lens galaxies, these other results are based primarily on the halo and stellar mass function constraints, and subject to a particular parameterisation of the SHMR.
While the uncertainties are large---but especially given the very different approaches to measuring these quantities \citep[see also][]{Dvornik2020}---it is encouraging that our results are at least broadly consistent, to within a factor of 2.5.

\subsubsection{Quantifying the relation between stellar mass and halo mass by fitting to the ensemble}

If we wanted to quantify the relation between stellar and halo mass using our
data, we could imagine fitting a linear relation to any one of the different
sets of binned and stacked halo mass measurements shown in \figref{logmstacks}.
In practice, however, the answer we would get would depend somewhat on what
binning we chose to adopt. This problem is compounded by the fact that the
uncertainties on each point is manifestly non-Gaussian, so the error
propagation would be non-trivial. It is for these reasons that we do not base
our analysis on the binned and stacked ESD profiles, but instead perform
simultaneous fits to the many independent ESD profiles for the lenses in our
sample, as described in \secref{fitting}. 

The results of this fit are shown in \figref{afologm}: we find ${\partial M\halo / \partial \log M_*} = 0.36^{+0.83}_{-0.93}$. 
In light of the systematic difference in the mean halo mass mentioned above, the comparison to past results is better done in terms of the logarithmic slope, ${\partial \log M\halo / \partial \log M_*} = 0.32^{+0.99}_{-0.90}$; the comparable 
values in this mass range are 0.71 and 0.72 for \citet{Moster2010} and \citet{vanUitert2016}, respectively.
In other words, we do infer a slightly shallower SHMR than either of these authors, but this difference is not statistically significant.
Again, given the very different methods used to arrive at these values, and allowing that the uncertainties are large, we take it as encouraging that the results are at least broadly consistent.

\begin{figure*} \centering
\includegraphics[width=8.6cm]{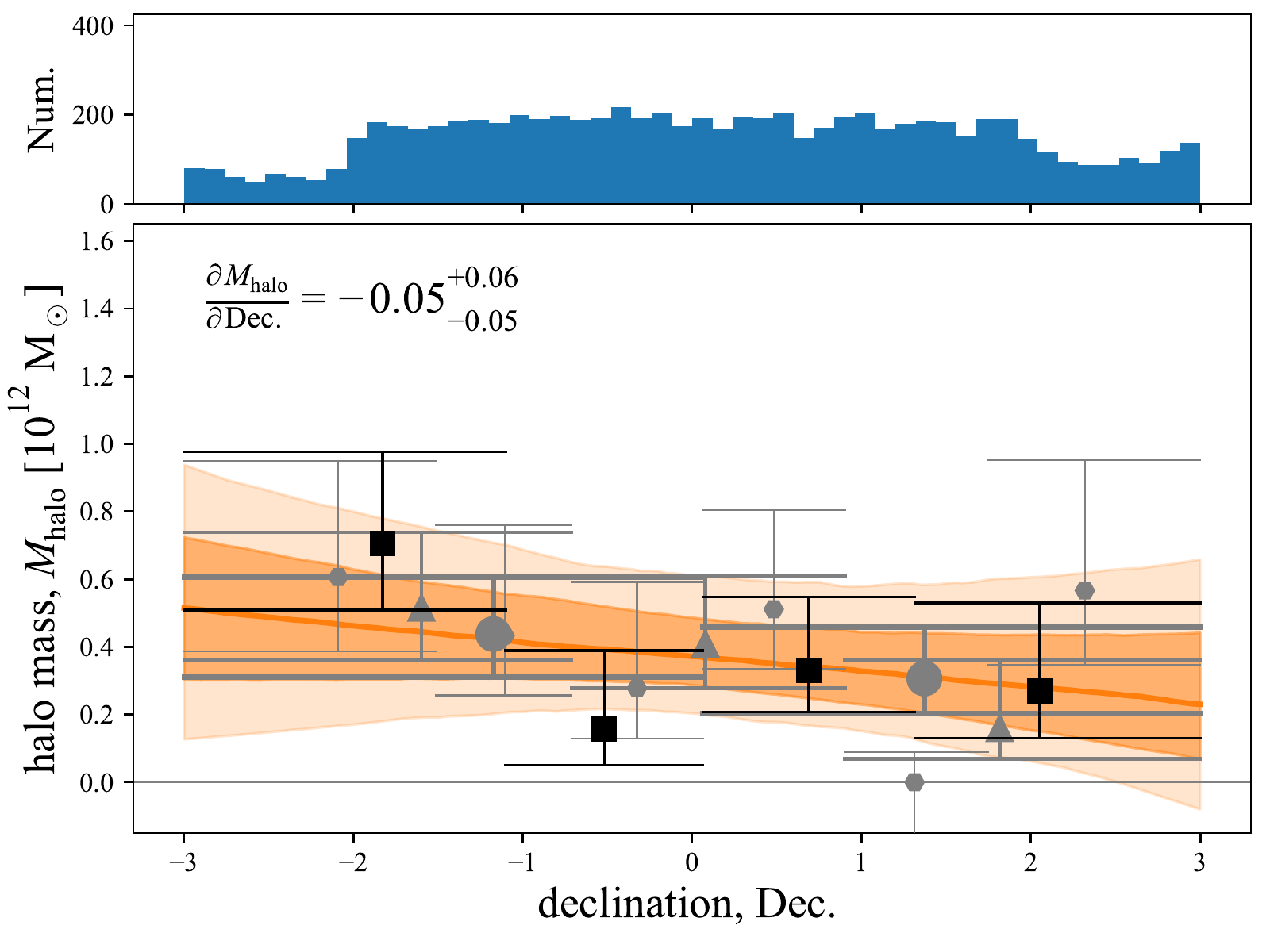}
\includegraphics[width=8.6cm]{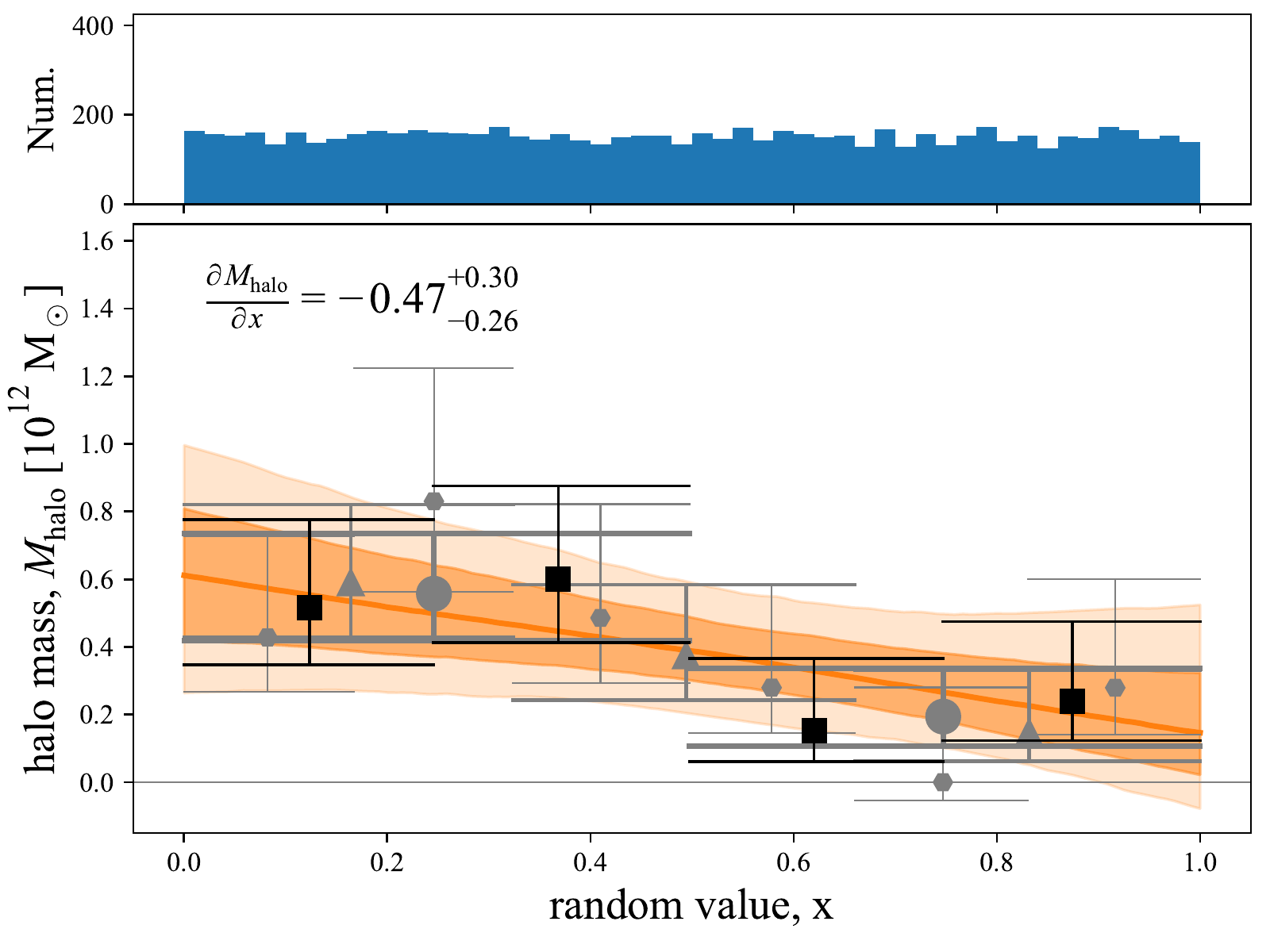}
\includegraphics[width=8.6cm]{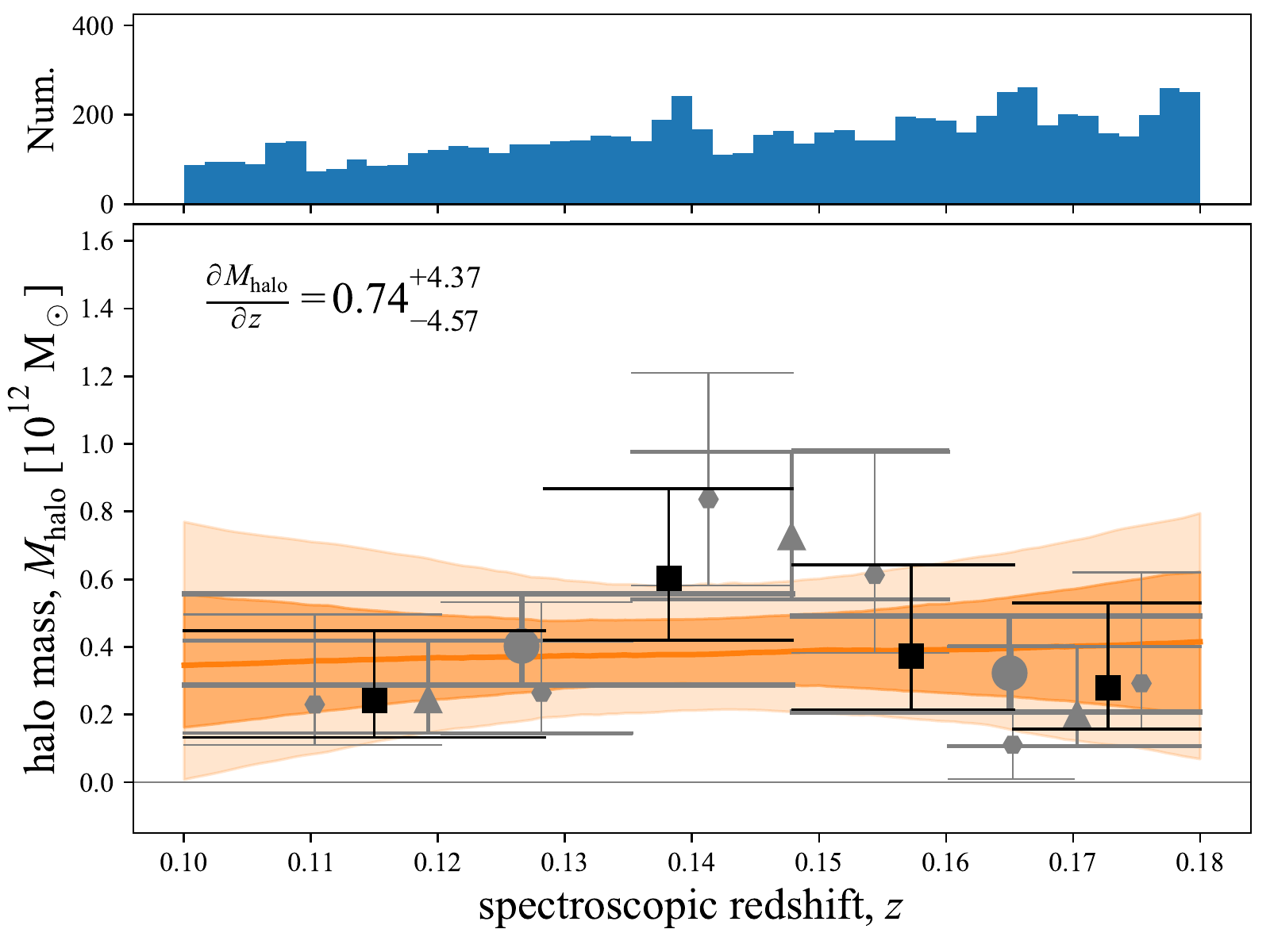}
\includegraphics[width=8.6cm]{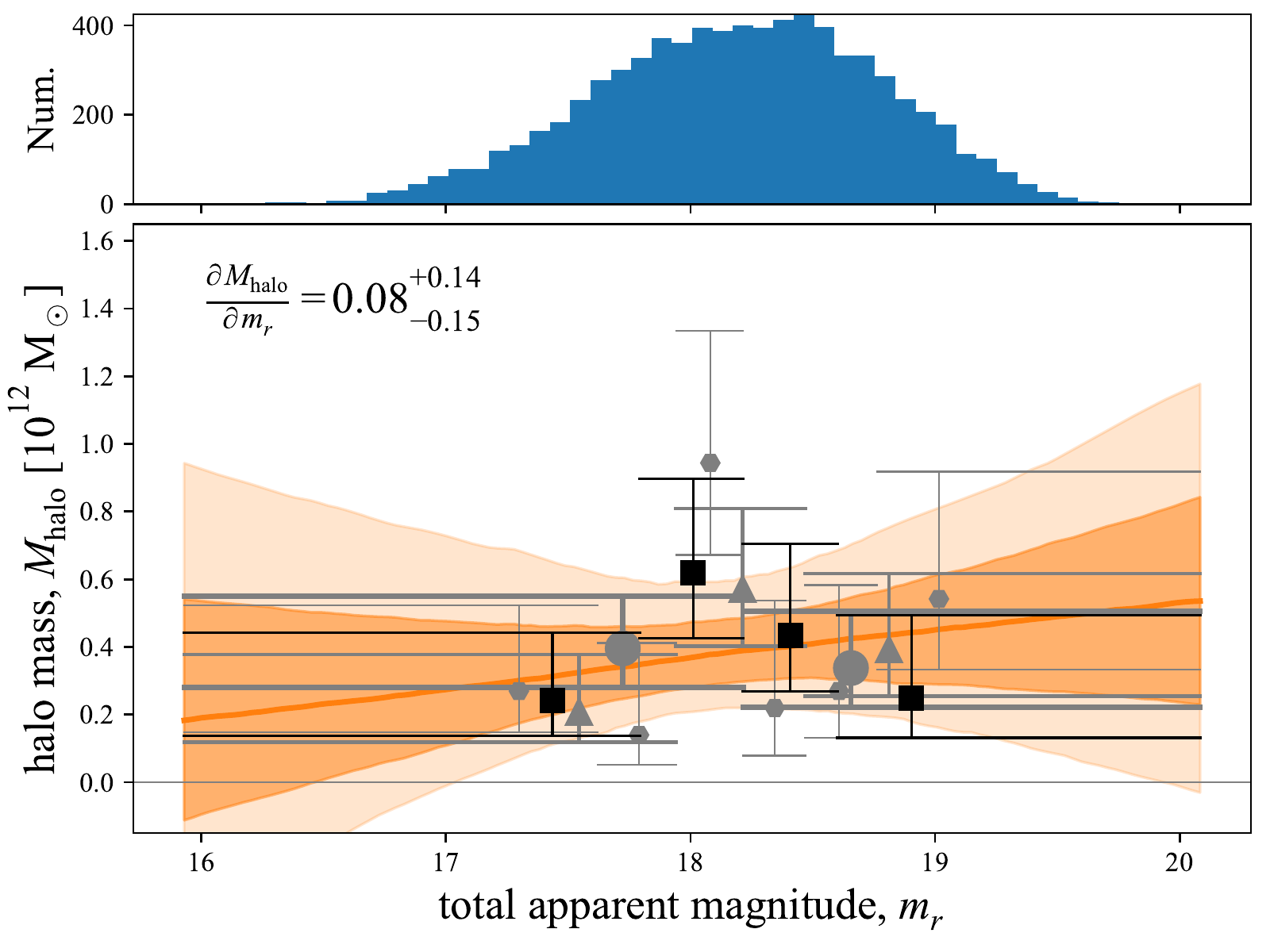}
\caption{Some null results as simple sanity checks for our analysis.--- Each plot shows the
apparent variation in halo mass as a function of some property that is expected
to be unrelated to halo mass. As in \figref{logmstacks} and \figref{afologm}, the histograms in the upper panels show the distributions of each property across our sample.
The fact that the observed dependence of halo
mass on each of these quantities is consistent with zero is reassuring. 
\label{fig:sanity}} \end{figure*}

\subsection{Some simple sanity checks, and the potential for bias} \label{sec:sanity}

Before we move on to searching for galaxy parameters that show statistical
correlations with halo mass, it is worthwhile to demonstrate a null result. In
\figref{sanity}, we show the inferred dependence of halo mass on several
parameters that we would expect to be completely unrelated to halo mass:
namely, declination, and a random value. We also show the inferred relation
between halo mass and both redshift and apparent magnitude. If there is no
significant evolution across our $0.10 < z < 0.18$ redshift window, and if our
sample is properly volume limited (\ie, that the stellar populations across our
sample do not vary with redshift), then we would expect to find ${\partial
M\halo / \partial z}$ = 0, and $\partial M\halo / \partial m_r = 0$. In
each of the four cases shown, while the uncertainties are large, the measured
values do conform to these simple expectations.

\figref{sanity}, and particularly the panel showing the inferred correlation
between $M\halo$ and apparent magnitude, is also useful to illustrate how our
scheme for allowing the value of $M\halo$ to be negative mitigates potential
biases in our results. As shown in this Figure, the data are sometimes
consistent with linear relations that would imply negative halo masses. Taken
at face value, negative masses would seem to be unphysical, but this is totally
consistent with a low signal-to-noise measurement of a strictly positive
quantity. \looseness-1

We have considered two alternatives to our preferred approach for accommodating
negative values for $M\halo$ (which is simply to define $\esd(-M\halo) \equiv
-\esd(M\halo)$; see \secref{fitting}). The first would be to restrict the
allowed region of $(A,~b)$ parameter space to forbid any non-positive values of
$M_{\mathrm{halo,i}}$. Looking at \figref{sanity}, the problem
with this approach becomes clear: the allowed range of fit parameters would
become very sensitive to the furthest outlying point and/or the precise limits
over which the fitting is done. Looking at the trend in halo mass as a function of
apparent magnitude, for example, we would get a very different answer if we
required $M\halo$ to be positive for $r < 17$, or 16, or 12. The net result
would be a potentially strong bias against large values of $A$ and/or small
values of $b$.  \looseness-1

\begin{figure*} \centering
\includegraphics[width=8.6cm]{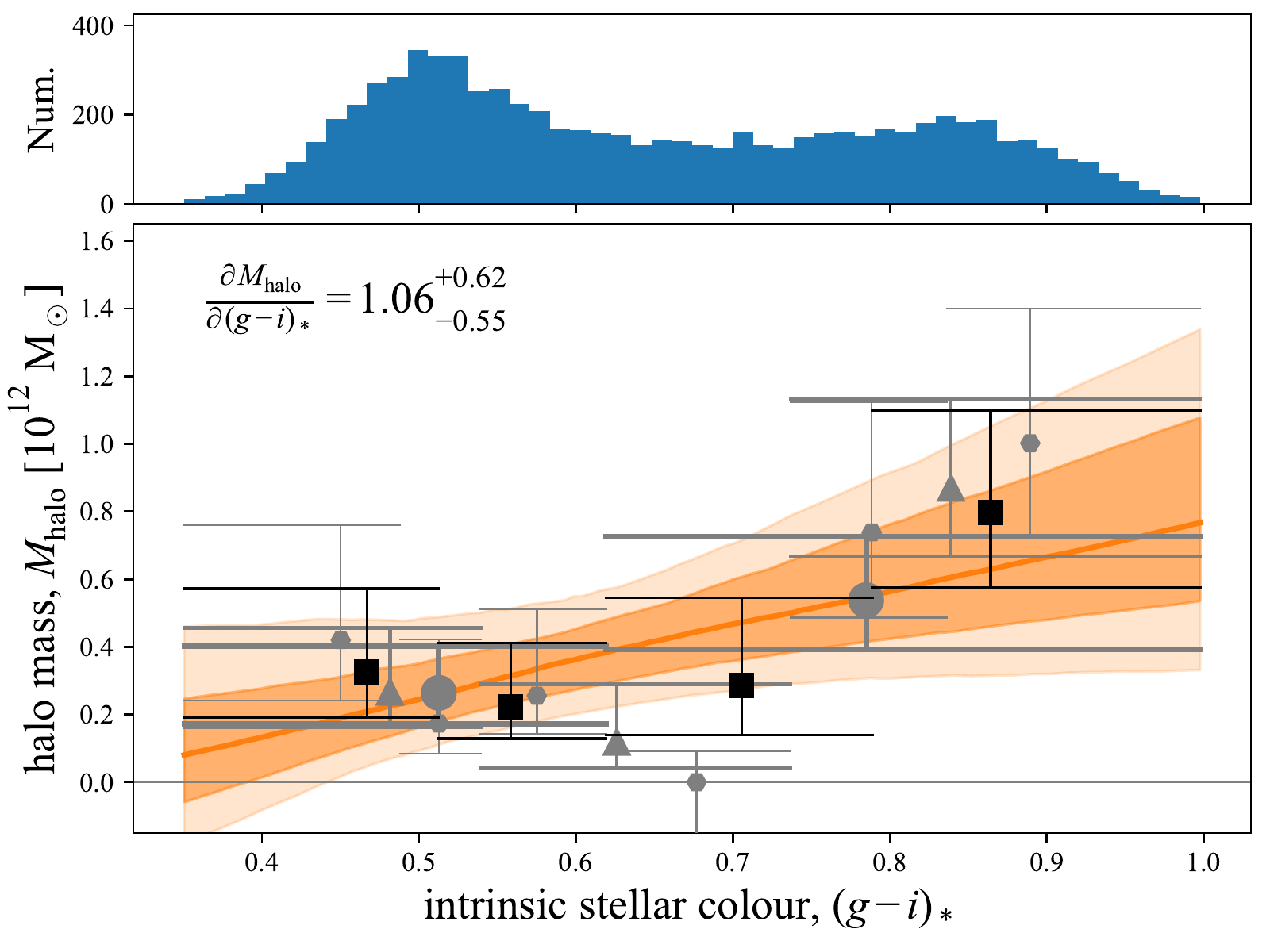}
\includegraphics[width=8.6cm]{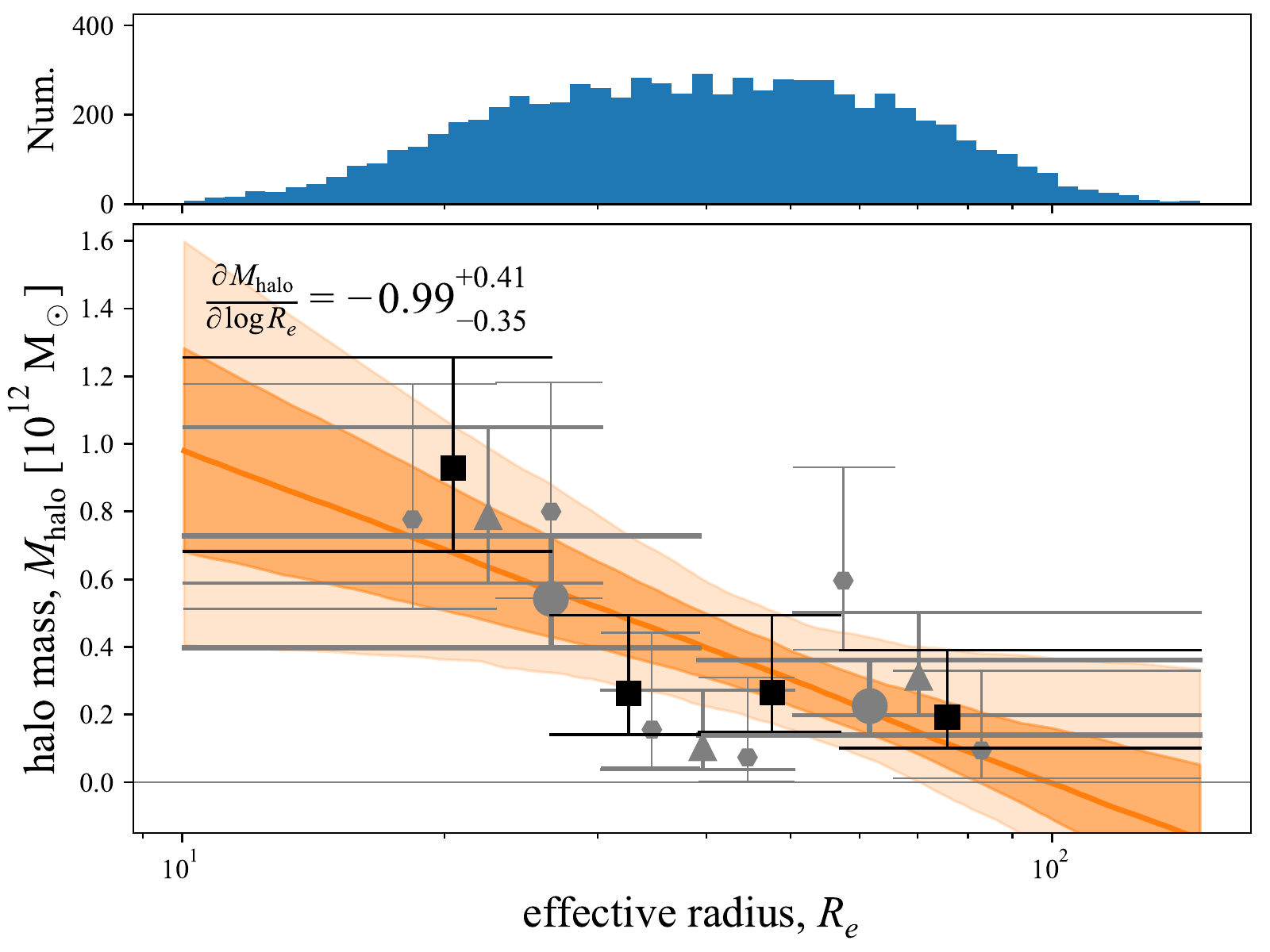}
\includegraphics[width=8.6cm]{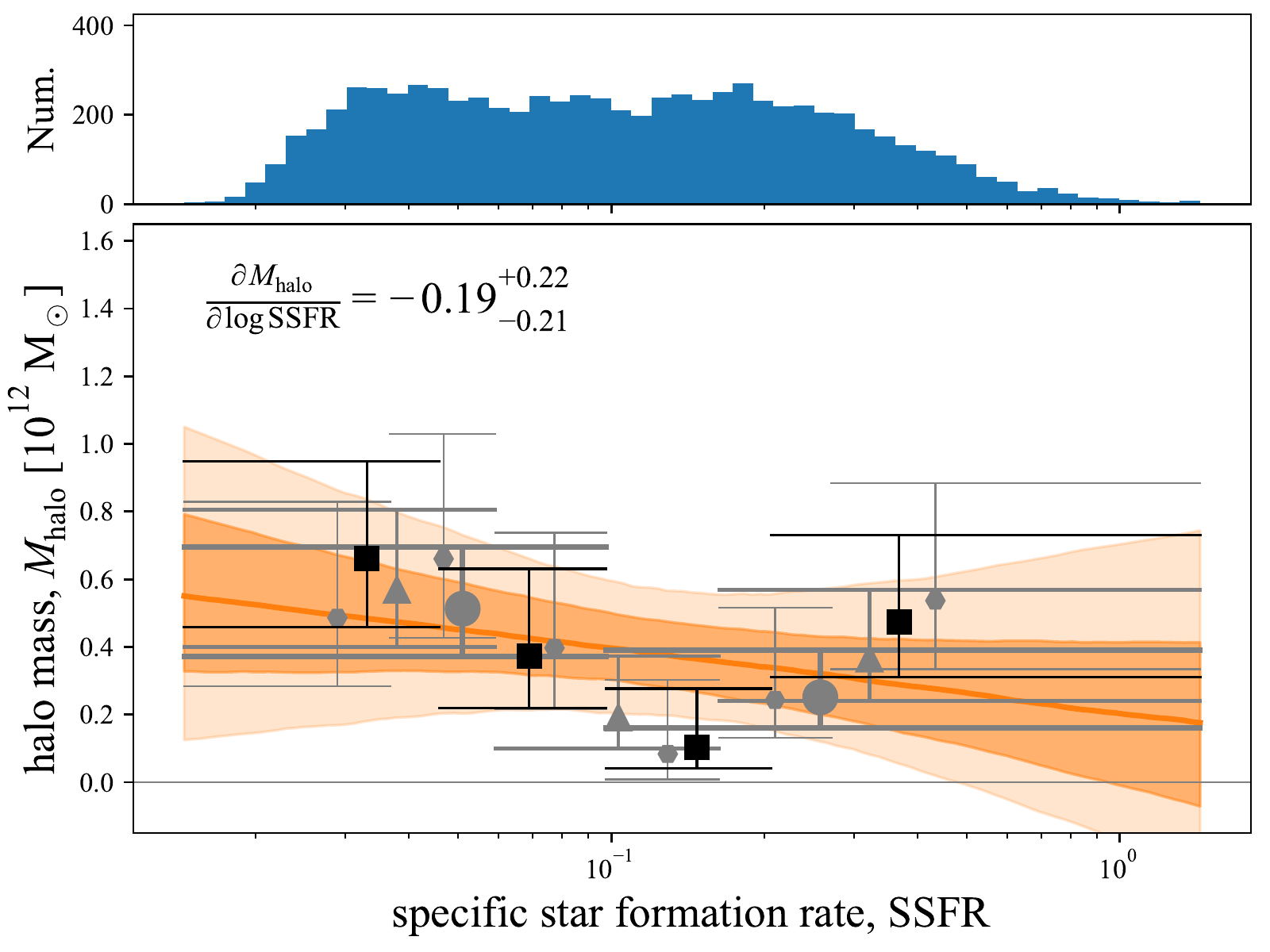}
\includegraphics[width=8.6cm]{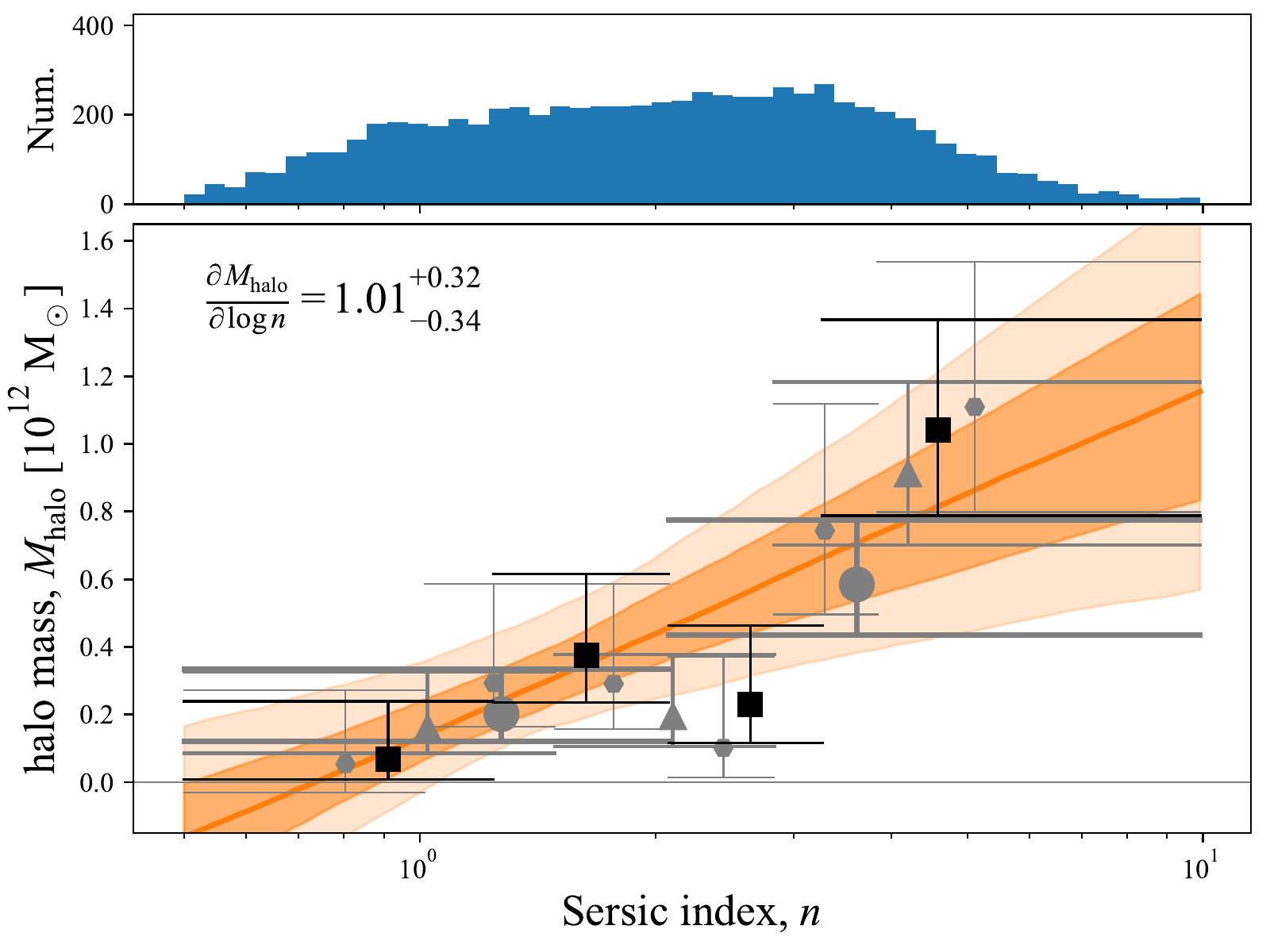}
\caption{Exploring variations in halo mass as a function of other key galaxy
observables.--- The different panels show the apparent trend in the mean halo
mass as a function of stellar colour, specific star formation rate, \Sersic\
index, and effective radius. Focusing on the points, which show the mean
inferred halo mass in bins, there is a clear and apparently linear correlation
between halo mass and both \Sersic\ index and effective radius, which is in close accord with the linear
fits to the sample as a whole. For stellar colour and SSFR, we do see
significant correlations with halo mass across the sample, but the binned
results are arguably less consistent with a simple linear dependence. 
\label{fig:trends}} \end{figure*}

At the other extreme, we could simply place a floor of zero on the halo mass values, or, equivalently, say that $\esd(M\halo) = 0$ for $M\halo \le 0$. 
With this decision, any observed ESD profile would be equally
well consistent with $M\halo = -10^{14}$ M\sol\ as 0, and only data in the
$M\halo > 0$ regime would contribute to the constraints on the values of $A$
and $b$. To the extent that approach would allow us to consider steeper
gradients, it would mean that we would be more likely to overestimate any
correlations than to underestimate them: that is, the net result would be a
potential bias towards larger values of $A$ and/or small values of $b$.

With these simple arguments, we motivate our specific approach to accommodating
negative values of $M\halo$ as being intermediate between these two extremes,
and with the hope that any bias in our results that arise as a consequence of
this decision are small.  \looseness-1

\section{Results: Exploring correlations between halo mass and global galaxy
parameters} \label{sec:results}

Our basic results are shown in \figref{trends}. These panels show the inferred
variation in halo mass as a function of several key galaxy properties:
intrinsic stellar colour, $\gistar$, which is a proxy for light-weighted mean
stellar age; specific star formation rate, SSFR; \Sersic\ shape parameter, $n$;
and half-light radius, $R\eff$. 
We note that we have also considered a number
of other galaxy properties, including effective colour (\ie, without correcting
for internal extinction), ellipticity, star formation rate, H$\alpha$
equivalent width, mass-to-light ratio, etc.  
While we do not show the results for all of the properties we have considered, we summarise the fit results in \tabref{summary}.
Our choice to focus our attention and discussion to these four parameters in particular is influenced by our belief that they are the more `fundamental' and/or robustly measured than other related quantities (\eg, preferring intrinsic stellar colour to effective colour, or SSFR to SFR or H$\alpha$ equivalent width, and preferring the bimodal but relatively flat distribution in $\gistar$ to the more strongly skewed distribution of $Dn_{4000}$), and because they are the parameters that show the strongest correlations with halo mass.

The panels in \figref{trends} are analogous and directly comparable to
\figref{afologm} and \figref{sanity}: the upper panels show the distribution
for each property across the full sample; the points show
the maximum likelihood values for the mean halo mass, in bins of the quantity in question; the horizontal
error bars show the width of each bin; the vertical error bars show the
(asymmetric) $\pm1\sigma$ confidence intervals for the inferred halo mass. Then, the lines show the inferred
correlation between halo mass and the quantity in question; the heavier and
lighter shaded regions show the equivalent of the 1 and 2$\sigma$ range for
these fits, as a function of the property in question; and the inferred values for
the slope of the linear fit are also given with 1$\sigma$ uncertainties in each
panel. 

Our first and most basic observation is that {\em even over the narrow range of stellar masses covered by our sample, there is significant variation in galaxies' halo masses}. 
Taking the stacked-by-quartile results in \figref{trends} at face value, the observed minimum-to-maximum variation in stellar-to-halo mass fractions across the sample is at least 1 dex (0.1--1.2 $\times 10^{12}$ M$_{\sun}$).

Further to this, {\em at (approximately) fixed stellar mass, there are clear correlations between global galaxy properties and halo mass}. More specifically, we see that canonically `early type' galaxies (\ie, red, quiescent, or elliptical galaxies) have larger halo masses than `late types' (\ie, blue, star forming, or disk galaxies). 
In general terms, this result is consistent with, \eg, \citet{Hoekstra2005}, \citet{Mandelbaum2006}, and others, who find offset SHMRs for generically red/early versus blue/late samples. Our results offer additional detail and insight, by beginning to map the relations between these properties and halo mass at fixed stellar mass.

The obvious next question is: which property or properties are most closely correlated with halo mass? This is the question that occupies the rest of this paper.

Considering first the empirical correlation between halo mass and intrinsic stellar colour, $\gistar$, the binned results appear broadly consistent with all blue ($\gistar \lesssim 0.75$) galaxies in the sample having $\mhalo \approx 2$--$4 \times 10^{11}$ M\sol, and red ($\gistar \gtrsim 0.075$) galaxies having $\mhalo \approx 8 \times 10^{11}$ M\sol. 

Let us entertain this scenario for a moment to tease out an important aspect of our approach.
While there is nothing preventing us from fitting a linear relation between any two properties (\eg, to quantify the strength of the correlation), it should also be recognised that there is no guarantee that the result of such a fit will provide a good description of reality.
In order to have confidence that the linear fit provides a faithful description of the underlying data, therefore, what we are looking for is consistency between the the fits and the binned results. 
Conversely, where the binned results are not consistent with the linear fits, this is a sign that the relationship between these two properties is more complex than a simple linear correlation.
That is why we show both the binned values as well as the linear fits: as complementary representations of the same underlying data.
At the same time, recognising the size of the formal uncertainties in these panels, we must acknowledge the real risk of over-interpreting the data in this regard. 

Turning next at the empirical correlation between halo mass and effective radius, $R\eff$, the indication is that galaxies with $R\eff \gtrsim 4$ kpc (\ie, the canonical `late types', which have larger sizes, bluer colours, and lower \Sersic\ indices) have approximately constant halo masses $\approx 3$--$5 \times 10^{11}$ M\sol. 
There is possibly the suggestion from the binned results that there is a strong correlation between effective radius and halo mass for canonical `early types', such that more compact galaxies have higher halo masses, but for the reasons above, this cannot be taken as anything more than suggestive.  

Parenthetically, we note that the inverse correlation between $M\halo$ and $R\eff$ seen in \figref{trends} would seem, on its face, to be counter to the results in \citet{Charlton2017}, who find a positive correlation between the offsets from the SHMR and size--mass relations: averaging over $9 \lesssim \log M_* \lesssim 11.5$ and $0.2 < z_\mathrm{phot} < 0.8$, their result is $\Delta \log M\halo = (0.42 \pm 0.16) \, \Delta \log R\eff$. Noting the major differences in how our results are derived (volume limited vs.\ magnitude selected lens samples; spectroscopic vs.\ photometric redshifts used for lens selection and characterisation; narrow vs.\ broad redshift windows; exclusion vs.\ inclusion of satellites), we cannot hope to uniquely identify the cause for this apparent tension. That said, we do highlight that the \citet{Charlton2017} results are derived in bins of absolute magnitude {\em and} effective colour, with both offsets determined separately for the red and for blue galaxies subsamples. In this sense, the \citet{Charlton2017} results are perhaps better viewed as probing {\em third}-order correlations between $M\halo$ and $R\eff$ at fixed mass and colour --- and assuming that colour is the correct choice for the second-order term. 

Returning to the main discussion, the empirical correlation between halo mass and specific star formation rate is less impressive than the other parameters we consider. 
There is maybe a hint that there is a correlation
between halo mass and SSFR for actively star forming galaxies, and maybe also a
slight inverse correlation for more quiescent galaxies. The fact that the
variation in halo mass seen in this panel is less than in some others imply
that SSFR does not play a primary role in predicting halo mass (and vice
versa). Instead, we judge it more likely that the empirical correlation seen is
a spurious correlation induced by correlation between both SSFR and halo mass
and some other parameter(s).\footnote{This view is informed by a series of
numerical experiments where we explore the potential for `spurious'
correlations induced by more `fundamental' correlation between one particular
parameter and halo mass. More specifically, we find that the empirical
correlation between halo mass and SSFR can be largely or fully explained as a
natural consequence of the empirical correlation between halo mass and any of
\Sersic\ index, effective radius, or intrinsic stellar colour. The
converse---that a correlation between halo mass and SSFR can explain the
observed correlations with other properties---is untrue.}

Turning finally to the empirical correlation between \Sersic\ index and halo mass, here the binned data are most consistent with a simple linear correlation between $\log n$ and halo mass. 
This is also the most significant (in a statistical sense) correlation that we see in our dataset.

In summary, then, we have three possible ways of understanding our data: 
\begin{enumerate}
    \item a direct correlation between halo mass and galaxy concentration, as quantified by \Sersic\ index, $n$.
    \item a correlation between halo mass and size for early type galaxies, with late types having approximately constant halo masses; or
    \item a bimodal distribution of halo masses, which is tied to the bimodality in intrinsic stellar colour, $\gistar$.
\end{enumerate}
Or, of course, any combination of these three. 
While we cannot unambiguously discriminate between these scenarios with the present data, we discuss potential interpretations of these results further in the next section.

\begin{table*}
\begin{tabular}{ l r c r c c }
\hline\hline
quantity, $x$
& mean value, $x_0$
& $b = \langle M_\mathrm{halo} \rangle$
& $A = {\partial M_\mathrm{halo} / \partial x}$
& RMS value, $\sigma_X$
& implied SHMR dispersion 
        \\ \hline
{\em{\hspace{1.2cm}Principal quantities of interest: } } &&&& \\
specific star formation rate, $\log \mathrm{SSFR}$ / [Gyr$^{-1}$] & $-1.0103$ & $0.398_{-0.094}^{+0.106}$ & $-0.188_{-0.214}^{+0.223}$ & 0.400 & $0.09_{-0.06}^{+0.08}$ \\
intrinsic stellar colour, $(g-i)_*$ & $0.6192$ & $0.384_{-0.093}^{+0.095}$ & $1.065_{-0.549}^{+0.623}$ & 0.158 & $0.20_{-0.10}^{+0.10}$ \\ 
effective radius, $\log R_e$ / [kpc]  & $1.5942$ & $0.406_{-0.091}^{+0.101}$ & $-0.989_{-0.352}^{+0.411}$ & 0.226 & $0.24_{-0.09}^{+0.06}$ \\ 
\Sersic\ index, $\log n$ & $0.3179$ & $0.457_{-0.095}^{+0.122}$ & $1.007_{-0.335}^{+0.320}$ & 0.287 & $0.28_{-0.07}^{+0.06}$ \\ 

{\em{\hspace{1.2cm}Other astrophysical quantities: } } &&&& \\

stellar mass, $\log M_*$ / [M\sol] & $10.4865$ & $0.388_{-0.094}^{+0.105}$ & $0.357_{-0.928}^{+0.834}$ & 0.115 & $0.09_{-0.06}^{+0.09}$ \\ 
star formation rate,  $\log \mathrm{SFR}$  [M\sol\ yr$^{-1}$] & $0.4616$ & $0.407_{-0.105}^{+0.100}$ & $-0.153_{-0.230}^{+0.244}$ & 0.392 & $0.08_{-0.06}^{+0.08}$ \\ 
effective stellar colour, $(g-i)$ & $1.0449$ & $0.401_{-0.098}^{+0.104}$ & $0.388_{-0.564}^{+0.517}$ & 0.145 & $0.08_{-0.05}^{+0.06}$ \\ 
4000 \AA\ break strength,  $Dn_{4000}$ & $1.4838$ & $0.335_{-0.093}^{+0.095}$ & $0.629_{-0.310}^{+0.291}$ & 0.279 & $0.23_{-0.11}^{+0.11}$ \\ 

{\em{\hspace{1.2cm}Null results and controls: } } &&&& \\

\Sersic\ magnitude (apparent),  $m_r$ & $18.2147$ & $0.389_{-0.092}^{+0.096}$ & $0.084_{-0.154}^{+0.136}$ & 0.462 & $0.06_{-0.04}^{+0.06}$ \\ 
redshift,  $z$ & $0.1479$ & $0.387_{-0.101}^{+0.107}$ & $0.744_{-4.575}^{+4.367}$ & 0.022 & $0.08_{-0.06}^{+0.07}$ \\ 
Declination,  $\mathrm{Dec.}$ / [deg.]\ & $0.0631$ & $0.370_{-0.086}^{+0.113}$ & $-0.048_{-0.050}^{+0.061}$ & 1.502 & $0.10_{-0.07}^{+0.08}$ \\ 
random value, $x$ & $0.4973$ & $0.388_{-0.086}^{+0.101}$ & $-0.467_{-0.265}^{+0.300}$ & 0.289 & $0.16_{-0.09}^{+0.08}$ \\
axis ratio,  $q$ & $0.4105$ & $0.409_{-0.100}^{+0.095}$ & $-0.466_{-0.424}^{+0.457}$ & 0.224 & $0.12_{-0.08}^{+0.10}$ \\ 
position angle, $\theta$ / [deg] & $0.0813$ & $0.406_{-0.104}^{+0.101}$ & $-0.003_{-0.002}^{+0.002}$ & 51.485 & $0.19_{-0.10}^{+0.10}$ \\ 
    \hline \hline
\end{tabular}
\caption{Summarising the results of our linear fits to the correlation between halo mass and various quantities. --- For each quantity, $x$, the first three columns give the values for the linear relation $M\halo = A \, (x - x_0) + b$, as defined in \eqref{mhi}. As described in \secref{scatter}, we use $\ln 10 \, \sigma_x \, A / b$ as a metric to compare the relative strength of the correlation between each property and halo mass.  
These values, given in the final column, can be interpreted as the amount of dispersion in the SHMR that is directly coupled to the property in question, and as such, they provide an approximate lower bound on the dispersion around the SHMR. \label{tab:summary}}
\end{table*}

\section{Discussion: Probing the apparent connection between galaxy properties
and halo mass}

\begin{figure*}
\includegraphics[width=8.6cm]{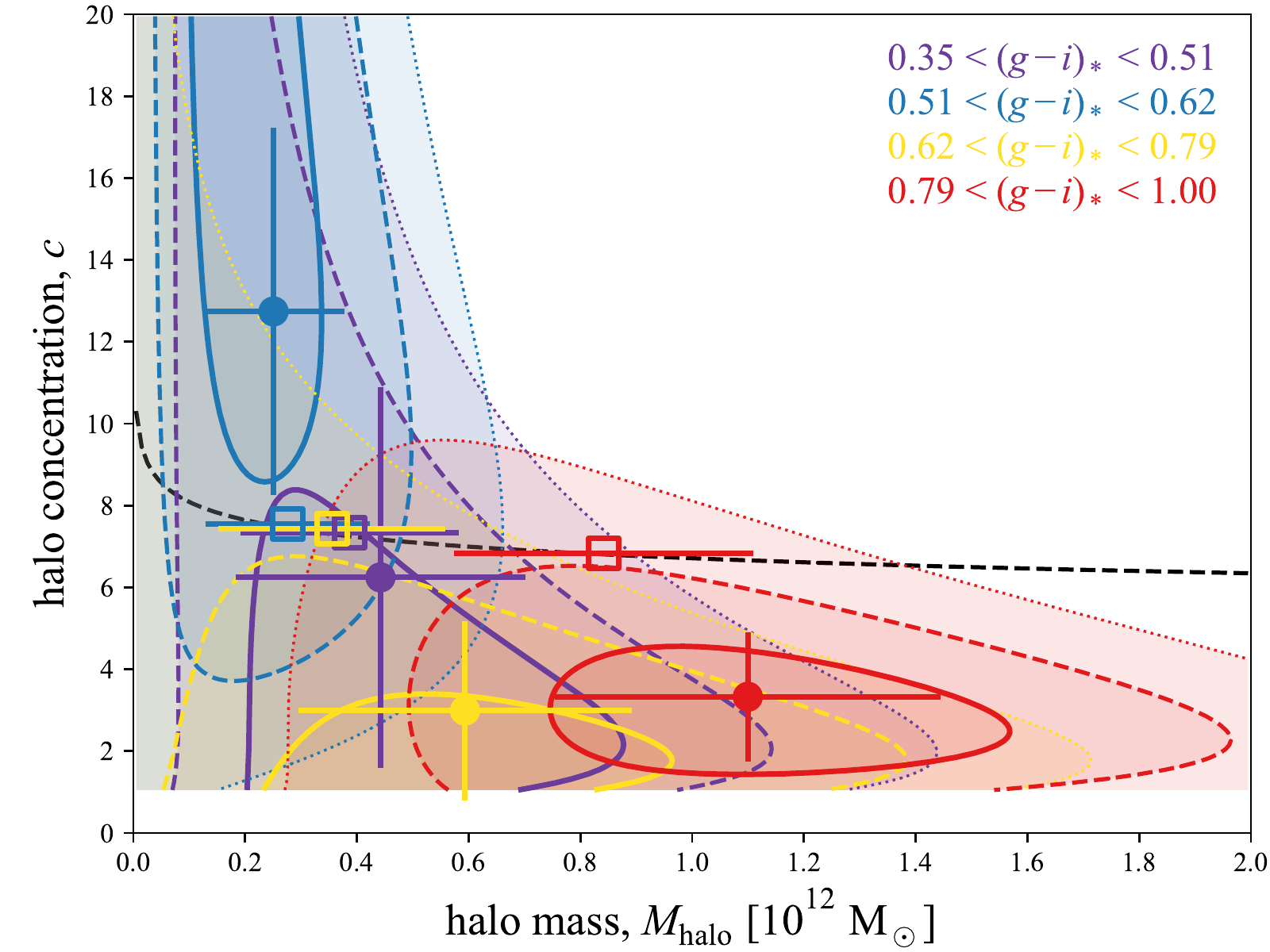}
\includegraphics[width=8.6cm]{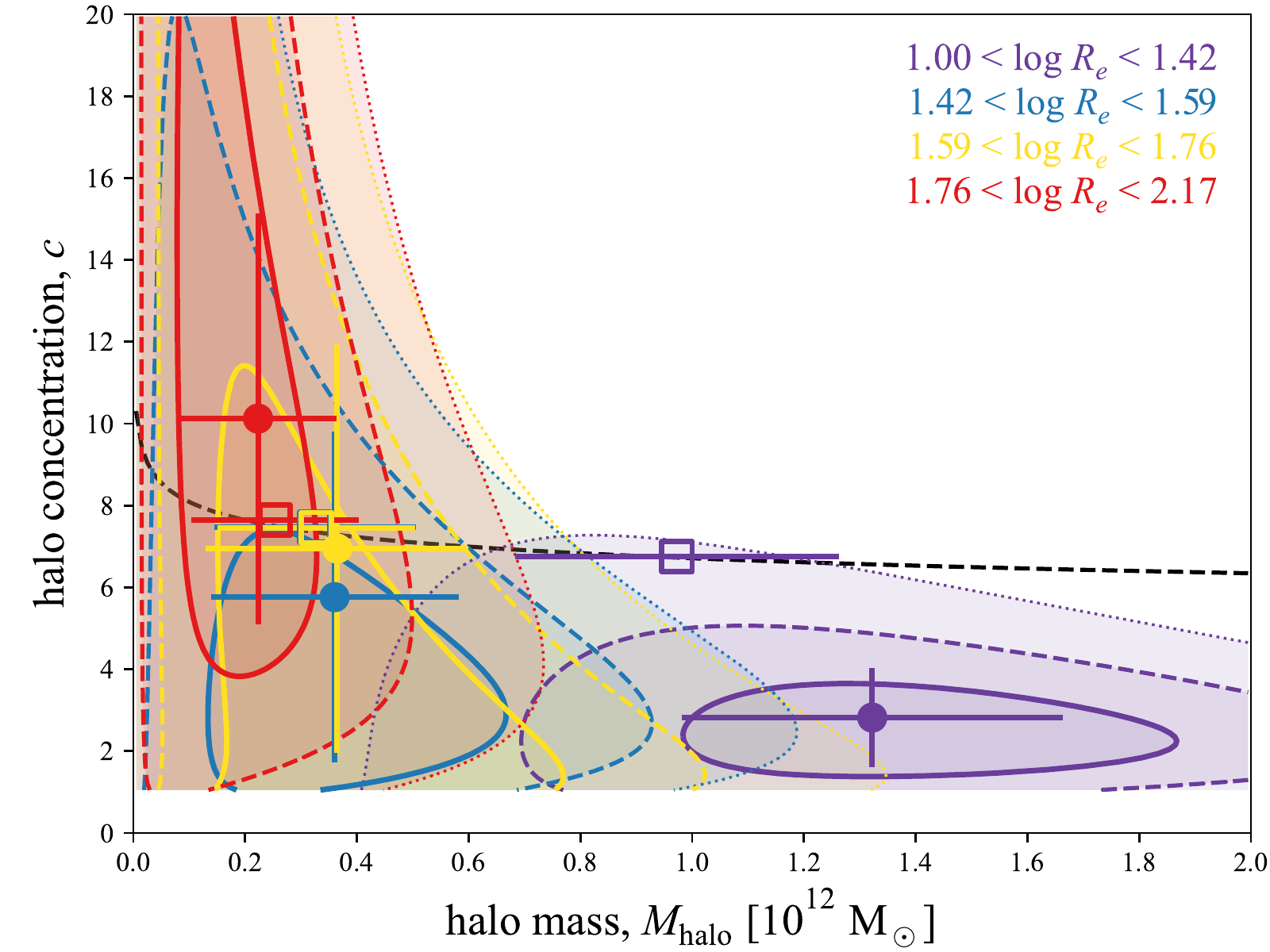}
\includegraphics[width=8.6cm]{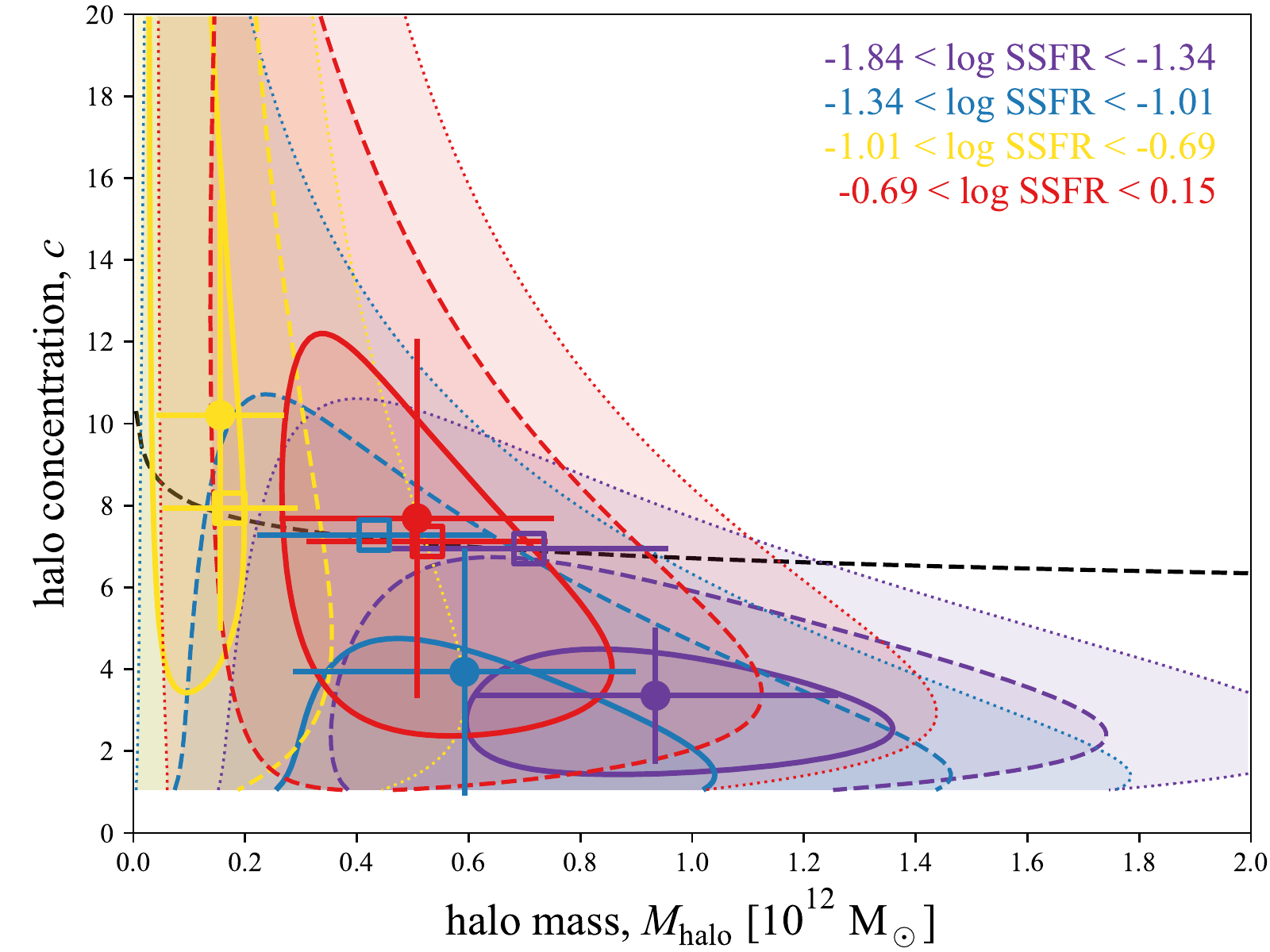}
\includegraphics[width=8.6cm]{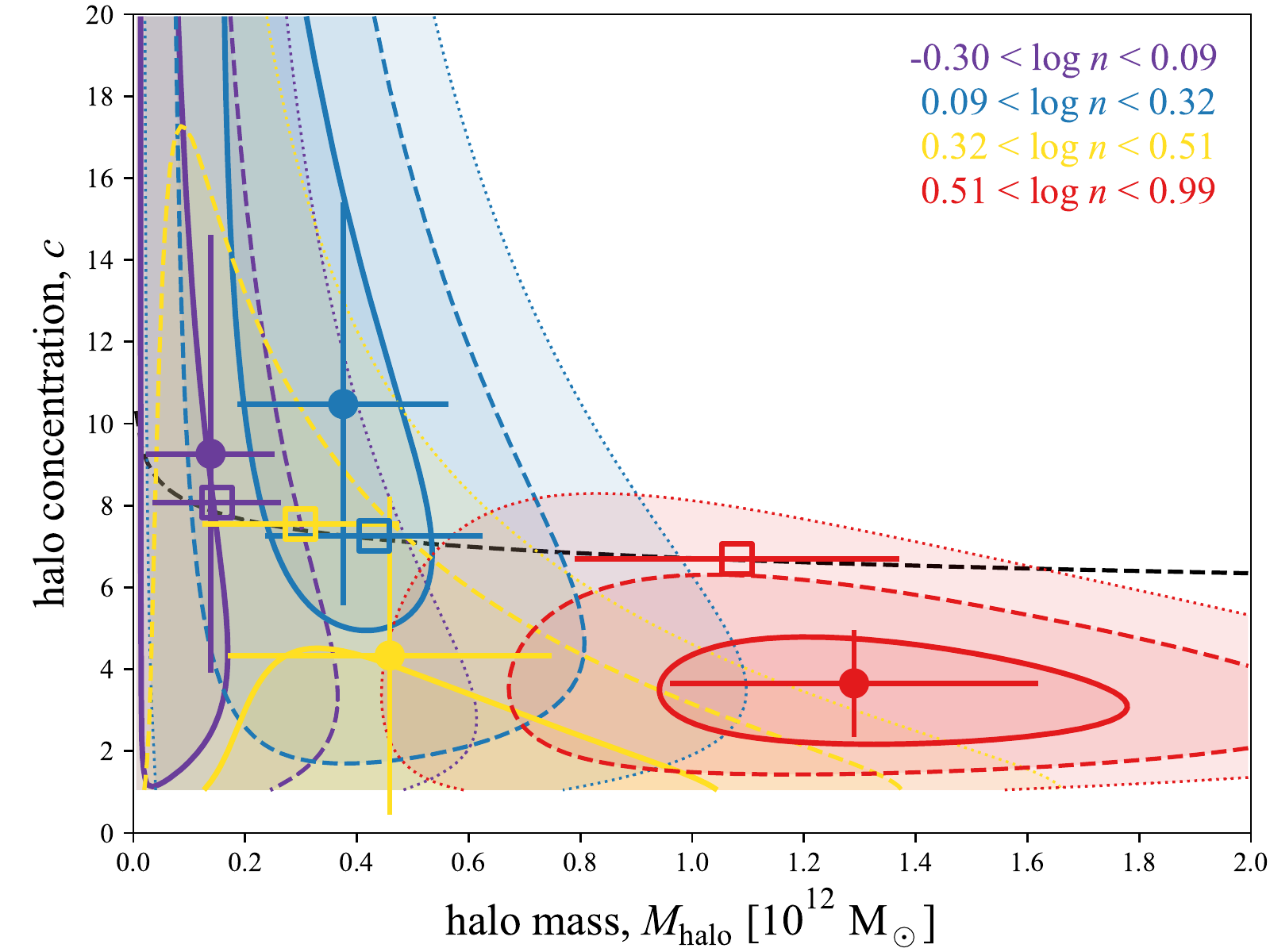}
    \caption{The joint constraints on halo mass and halo concentration, when binning
    by stellar colour, SSFR, effective radius, and \Sersic\ index. --- These
    figures provide a means to test the proposition that it is really halo mass
    that is varying across the sample, rather than halo concentration. The solid,
    dashed, and dotted contours in each panel show the region bounding the
    50\%, 90\%, and 99\% confidence regions in the halo mass-concentration
    plane when fitting to stacked ESD profiles for the sample split according
    to different properties. The filled points with error bars show the usual
    least-squares estimator and (symmetric) standard error estimates, which are what one would
    obtain by marginalising over either $M\halo$ or $c$. The black dashed
    line shows the \citet{Duffy2008} $M\halo$--$c$ relation that we adopt
    as our prior elsewhere in the text; the open squares show the values of
    $M\halo$ we obtain with this prior. In each case, the results for each
    subsample are consistent with no systematic variation in halo concentration
    across the sample as a whole, where there is clear evidence for significant
    variation in halo mass. This shows that we really are seeing variations in
    halo mass, rather than halo shape, across our sample. \label{fig:massconc}}
\end{figure*}

\subsection{Quantifying the connection between halo mass and galaxy properties}
\label{sec:scatter} 

The essential idea that we explore in this section is the extent to which, at fixed mass, the scatter around the mean/median stellar-to-halo mass relation (SHMR) is directly coupled to one (and only one) particular galaxy property.
Our motivation here is that by finding the galaxy property that is most directly associated with halo mass, we then have circumstantial evidence for the primary mechanism(s) by which halo mass influences galaxy evolution. 
Such a finding would also have implications for, \eg, cosmological studies where assembly bias is relevant.

Given the tight correlations between many different galaxy observables, how could we hope to identify such a property? 
The simplest approach---and the only one that we might hope our data to support---is to find the property that implies the largest spread of halo masses across the sample. 
To see this, imagine that there is a tight correlation between halo mass and some galaxy property, $x$, and that this property $x$ is itself correlated with some other property, $y$, but with some scatter. 
To the extent that binning by $y$ results in mixing galaxies with different values for $x$---and by extension, $M\halo$---in each bin, the effect will be to drive the mean value of $M\halo$ in each bin towards the average value for the sample as a whole. 
(Think of the central limit theorem.)
The result is, therefore, that the observed (mean) trend in halo mass as a function of $y$ {\em will be less} than that as a function of $x$.\footnote{
By the same token, we cannot exclude the possibility that any (or even all) of the four apparent correlations shown in \figref{trends} is `spurious', in the sense that they are simply a consequence of a properly `fundamental' astrophysical correlation between halo mass and some property, unidentified, which is itself correlated with each the observables we consider---but this is inescapable.}

Another way of framing this question is: what galaxy property is the best predictor of halo mass across our sample
of $\log M_* \approx 10.5$ galaxies? 
By writing
\(  M\halo = A (x - x_0) + b + \epsilon  \)
where $\epsilon$ represents the offset from the mean relation between $M\halo$ and $x$, we can see the net dispersion in the SHMR as the sum (in quadrature) of two terms.
The first term is tied directly to the gross trend between $M\halo$ and $x$: \viz, $A \sigma_x$, where $\sigma_x$ represents the distribution of $x$ across the sample via the RMS value.
The second term represents some `random' or unknown dispersion around that mean relation via the RMS value of the (unknown) $\epsilon$s.
Since the net dispersion is a finite quantity determined by astrophysics, it can be considered fixed.
The best predictor of $M\halo$ is the quantity for which the $\epsilon$s are minimised, or, conversely, where the value of $A \sigma_x$ is maximised.
The quantity $A \sigma_x$ can also be seen as providing an approximate lower
limit on the true dispersion in the SHMR, inasmuch as any intrinsic scatter around the relation between $M\halo$ and $x$ will propagate through to a greater net dispersion.

With this in mind, we use the implied dispersion in the SHMR as a quantitative basis for comparing the strength or significance of the correlation between halo mass and other observables. In our motivating remarks above, which assume Gaussian statistics, we have used $M\halo$.  Since the SHMR dispersion is expected to be lognormal, it makes more astrophysical sense to consider the Gaussian dispersion in $\log M\halo$; \ie:
\begin{equation} \label{eq:scatter}
\sigma_{\log M\halo} 
    \gtrsim {\partial \log M\halo \over \partial x} \, \sigma_x 
    = \ln 10 \, {\partial M\halo \over \partial x }
            \, { \sigma_x \over \langle M\halo \rangle } 
                ~ ,
\end{equation}
where values for ${\partial M\halo / \partial x } = A$ and $\langle M\halo \rangle = b$ come directly from the MCMC chains for our modelling.
We note that since this quantity is pure scalar (\ie, is dimensionless), it is robust to any gross systematic errors in our halo mass measurements that might arise due to choice of cosmology, halo modelling, shape or redshift measurement systematics, etc.
For this reason, we strongly recommend the use of this parameter for comparisons between our observational results and the results of other studies or models.

The inferred values for the SHMR dispersion, derived in this way, are given in
\tabref{summary} for all of the properties we have considered. 
Taking these estimated dispersions at face value, we would conclude that the
property most directly related to halo mass is \Sersic\ index, with an implied
SHMR dispersion $\gtrsim 0.28$ dex. The strength of the correlations between
halo mass and  effective radius and intrinsic stellar colour are only slightly
weaker, with an implied SHMR dispersion $\gtrsim 0.24$ dex and $\gtrsim 0.20$
dex, respectively. 

\subsection{Is it really not just stellar mass?} \label{sec:masserrors}

Our analysis specifically focuses on a narrow range of stellar mass, in an
attempt to identify correlations with halo mass at fixed stellar mass. The
formal random errors in the stellar mass estimates that our selection is based
on are typically 0.12 dex, which is not negligible in comparison to our 0.4 dex selection window. 
The particular concern here would come
from Eddington-like biases: while individual galaxies are as likely to scatter
to higher or lower masses, the fact that lower mass galaxies are more common
produces a systematic bias as more low mass galaxies scatter up into the
sample than high mass galaxies scatter down. The consequence of this would be
that we might be overestimating stellar masses nearer to our lower mass limit,
and underestimating those nearer to our upper mass limit. Simple numerical
experiments suggest that despite the fact that $\sim 20$ \% more galaxies
scatter across the lower selection boundary than the upper one, the scale of
this kind of bias on the mean mass is $< 0.006$ dex at the low mass end, and
even smaller at the high mass end. In short, the systematic impact that random errors in
the stellar mass estimates might have on our results really is negligible. \looseness-1

The bigger concern might be differential systematic errors in the stellar mass
estimates, whereby there is some bias in galaxies' stellar mass estimates that
correlates directly with one or more of the observable properties we have
looked at. While we cannot unambiguously rule out this possibility, we can ask
how big such a bias would have to be in order to fully explain our results,
following a simple argument based on propagation of errors. In order to explain
the apparent correlation between halo mass and \Sersic\ index, the size of the
differential systematic bias would have to be such that we are underestimating
the stellar masses of high $n$ galaxies by $\sim 0.4$ dex (\ie, a factor of $\approx$
2.5) relative to low $n$ galaxies; this would be equivalent to missing 1
magnitude of flux. For size, the bias would have to be such that we are
overestimating the masses of large galaxies by $\approx 0.25$ dex, or a factor
of $\approx$ 1.8, relative to small galaxies. For colour, the masses of red
galaxies would have to be underestimated by $\approx 0.35$ dex, or a factor of
$\sim 2.25$, relative to blue galaxies. While biases of this size are not
inconceivable
\citep[see][for constraints on differential systematic errors on stellar mass estimates]{Taylor2010}, they would certainly be extreme, to the point of undermining the vast majority
of work on galaxy formation and evolution over the past several decades.

\subsection{Is it really halo mass?} \label{sec:concmass}

At the most basic level,
what we have shown is that there are differences in the observed ESD profiles
for lens samples split by different lens galaxy properties, which we have interpreted  as being due to variations in mean
halo mass across the sample. Another possibility is that the observed
differences in the ESDs profile are in fact the result of variations of some
halo property other than mass; \eg, halo concentration. 

Recall from \secref{fitting} that we are forced to assume a strict prior
on halo concentration as a function of halo mass. This is because we cannot properly
constrain the values of either $c$ or $M\halo$ for our lens sample without such
a prior. In this section, we do away with this prior, and look at the joint or
bivariate halo mass--concentration constraints, $P(M_\mathrm{halo},c)$, that can be derived from our
data.

The change is just to allow $c$ to be a free parameter in the NFW description
of the halo ESD profile, $\Sigma_\mathrm{NFW}(R|M\halo, c)$ so that \eqref{chi}
and \eqref{ell} become bivariate expressions of both $M\halo$ and $c$, rather
than $M\halo$ alone. What we have done is to map the bivariate likelihood
function, $\Ell(M\halo,c)$, based on fits to the stacked ESD profiles for each
of our quartile subsamples.

The results of this exercise are shown in \figref{massconc}. 
While the uncertainties are large, there is not clear evidence for significant variation in halo concentration across the sample: the $c$ values for each subsample are generally consistent with one another. 
This is not true for halo mass, where there is clear evidence of
significant variations in the mean values for $M\halo$ across the sample. 
While we cannot be absolutely sure that it is mass that is uniquely driving the observed correlations, we can therefore exclude the possibility that our results are being driven {\em solely} by variations in halo concentration. \looseness-1

Further to this point, it is heartening that the data are broadly consistent with our assumed prior.  That said, there is also the possibility that the variation in $c$ as a function of $M_\mathrm{halo}$ across our sample (\ie, at approximately fixed stellar mass) is slightly steeper than the global relation for all halos. 
We also note that, in general, the inferred variation in $M_\mathrm{halo}$ across the sample would be {\em greater} if we were to relax our strong prior on $c$:
Looking at the subsample with the highest mean halo mass in each panel of \figref{massconc}, it can be seen that the inferred mean halo mass is $\sim 30$\% larger with a flat prior on $c$ compared to the much stronger $c(M)$ prior from \citet{Duffy2008}.  
While we have no objective basis from which to argue that one or the other value is more correct, what we can say is that a less restrictive prior on $c$ would only {\em increase} the apparent variations in halo mass across our sample, and so the inferred strength of the correlations between halo mass and galaxy properties.
This is another reason why our inferred values for the dispersion in the SHMR
should be viewed as lower limits. \looseness-1
    
\section{Summary and Conclusions} \label{sec:summary}

Our broad purpose with this paper has been to explore correlations between halo mass, as measured by galaxy-galaxy weak lensing, and other observable galaxy properties.
Our work is based on a volume-limited sample of central galaxies (see \figref{sample}) spanning
a narrow range in stellar mass ($10.3 < \log M_* < 10.7$) and redshift
($0.10 < z < 0.18$). This particular mass range is interesting because 1.)\ 
being near the knee in the galaxy stellar mass function
and the stellar-to-halo mass relation (SHMR), this represents the point of
transition between the low- and high-mass regimes of galaxy evolution; and 2.)\
being where the various bimodalities in galaxy properties are most pronounced, this is where there is the greatest spread of galaxy properties at fixed mass, and so where we have the greatest statistical lever arm for identifying correlations with halo mass. \looseness-1

From a technical standpoint, the first novel aspect of this work is that in addition to deriving halo mass measurements from stacked ESD profiles, we have also explicitly modelled the full set of un-stacked profiles for all of the lenses in our sample.  
Specifically, as described in \secref{fitting}, we have made linear fits to the relations between halo mass and galaxy observables, where we have considered all of the individual lenses in our sample simultaneously. 
This approach can be viewed as intermediate between convention and the two-dimensional modelling approach pursued in \citep{Dvornik2019, Dvornik2020}, which gets away from the stacking of many sources into azimuthally-averaged ESD profiles by considering each source independently.
We have validated our approach in \secref{proof} through comparison between our measurement of the SHMR over this narrow mass range and  previous results, and by demonstrating that we see no correlation between halo mass and a number of unrelated variables; \viz\ declination, redshift, or a random variable. 
We discuss the (limited) potential for systematic biases in our results in \secref{sanity}.

When considering the results shown in Fig.s \ref{fig:afologm}, \ref{fig:sanity}, and \ref{fig:trends}, it is important to recognise that the lines should not be understood as fits to the points. 
Where the points show the mean halo mass inferred from stacked ESD profiles, after binning by a particular galaxy property, the lines represent the inferred mean relation between halo mass and the property in question, as inferred from our modelling of the full ensemble of un-stacked ESD profiles. 
The points and the lines are therefore best understood as complementary representations of the general trend across the sample. 
Note that our analysis here is not directly sensitive to any additional scatter in halo mass around the mean relations as we observe them; see \secref{scatter} for further discussion of this point.

\figref{trends} shows our essential results, which are the empirical
correlations between halo mass, and several key galaxy properties; \viz\
intrinsic stellar colour (as an tracer of stellar populations generally light-weighted mean stellar age in particular), specific star formation rate, effective radius (as a proxy for size and/or density), and
\Sersic\ index (as an indicator of galaxy structure or concentration, which can also be taken as a proxy for bulge-to-disk ratio).
We see evidence for variation in halo mass as a function of each of the galaxy properties that we have considered.
In general terms, our main observational result is thus that, for the same stellar mass, canonically `early type' galaxies have larger halo masses than canonical `late types'. 

Our results are qualitiatively and quantitatively consistent with the stellar-to-halo mass measurements (in the relevant mass range)  of red versus blue galaxies by, \eg, \citet{Hoekstra2005}, \citet{Mandelbaum2006}, and \citet{Hudson2015}. 
As discussed in \secref{results}, there is at least superficial tension between our qualitative interpretation of the relation between halo mass and galaxy size and that of \citet{Charlton2017}.
Our results should be contrasted to complimentary studies by, \eg, \citet{Sonnenfeld2019} and \citet{Huang2020}, who have investigated stellar-to-halo mass ratios as a function of size/structure for very- to super-massive galaxies ($\log M_* > 11$ and $> 11.7$, respectively).  Our findings are also broadly consistent with those of \citet{AlpaslanTinker2020}, who show strong correlations between the properties of SDSS central galaxies and the combined luminosity of their satellites, which they use as a proxy for halo mass. \looseness-1

At a basic level, what we have shown is that there are differences in the lensing signal from halos  (more specifically, the ESD profiles; see \figref{esd_mstar}), which we are then interpreting as being due to variations in the mean halo masses, as a function of different observables. 
It is conceivable what we are seeing are really variations in some other halo property, which we are mistakenly attributing to variations in mass. 
An important factor in our process for inferring halo masses is a strong prior constraint on halo concentration as a function of mass, since we cannot place strong constraints on halo mass without such a prior. 
In order to address the possibility that what we may be seeing is variation in halo concentration, rather than mass, as a function of galaxy properties, we have looked at the joint halo mass--concentration likelihood surface for various subsamples of our data. 
As shown in \figref{massconc}, while the data are consistent with all subsamples having approximately the same halo concentration, there is clear evidence for variations in halo mass across the sample. 
It remains possible that there is also some variation in halo concentration across the sample, but it is clear that there is certainly significant variation in halo mass. 
It also seems likely that relaxing the $c(M\halo)$ prior would only strengthen the observed correlations, rather than reduce them.

As a way to derive new insights into the influence that halo mass has on the formation and evolution of individual galaxies, our particular goal is to identify galaxy properties that are most directly correlated with offset from the SHMR.
Saying the same thing in another way, we are looking for what galaxy property (in addition to stellar mass) is the best predictor of halo mass.
Compared to past studies, the novel aspects of this work are 1.)\ use of a narrow stellar mass range, to control for stellar mass dependence as best we can; and 2.)\ a systematic consideration of multiple galaxy observables. \looseness-1

The observed variation in mean halo mass as a function of galaxy properties within our sample demonstrates that $\log M_* \sim 10.5$ galaxies span a range of halo masses; that is, at fixed mass, there is significant dispersion in the SHMR. 
Moreover, the fact that the observed variations in halo mass across our sample as a function of colour, SSFR, size, and shape are larger than as a function of stellar mass clearly demonstrates that we are directly probing the dispersion in the SHMR. 
We can thus use our sample to get an approximate lower bound on the SHMR dispersion, under the assumption that the offset from the mean SHMR is fully and directly tied to one given observable (see \secref{scatter}).
In this way, we can limit the dispersion in the SHMR at $\log M_* \sim 10.5$ to be $\gtrsim 0.3$ dex. 

While the expectation from simulations is that there should be significant dispersion in the SHMR \citep[as high as $\sim 0.4$ dex;][]{Mitchell2016}, there are not yet many strong, direct observational constraints  \citep[but see, \eg,][]{Cao2019}.
Abundance matching and halo occupation modelling approaches typically find an inferred dispersion {\em in stellar mass at fixed halo mass} of order $0.2 \pm 0.02$ dex \citep[\eg][]{Moster2010,vanUitert2016,Tinker2017}.
This propagates through to an expected dispersion {\em in halo mass at fixed stellar mass} of order 0.24 dex for a $10.3 < \log M_* < 10.7$ sample like ours.
If there is any additional variation in stellar-to-halo mass ratios beyond what is directly correlated with galaxy structure, then the dispersion must be larger than the value of $\sim 0.2$ usually found by abundance/halo occupation modelling of the SHMR.
Alternatively, if past results are correct, then our results would suggest that the dispersion in the SHMR is essentially perfectly coupled to structure; in other words, that galaxies follow a sort of `fundamental plane' as a function of stellar mass, halo mass, and structure, with essentially no scatter. \looseness-1

It is not possible with the present dataset to definitively address the question of which parameter is (or parameters are) most directly and fundamentally tied to halo mass. 
But in general terms, we find that the structural properties of \Sersic\ index and effective radius are better predictors of halo mass than stellar population properties like stellar colour or star formation rate. 
The suggestion from the data is that \Sersic\ index is slightly preferred over effective radius as the `more fundamental' parameter: using the metric of the inferred dispersion in the SHMR to compare the relative significance of the trends with different parameters, the values are $0.28^{+0.06}_{-0.07}$ for \Sersic\ index and $0.24^{+0.06}_{-0.09}$ for effective radius, compared to $0.20^{+0.10}_{-0.10}$ for intrinsic stellar colour. 
The value for SSFR is just ${0.09^{+0.08}_{-0.06}}$, but with the caveats that 1.)\ the binned-and-stacked results indicate that a simple linear fit does not provide a faithful description of the mean relation between $M\halo$ and SSFR, and 2.)\ the observed trend with SSFR can be explained as a spurious or secondary correlation. \looseness-1

A naive interpretation of our results would be that, {\em at fixed stellar mass:} halo mass determines structure (but with some scatter); then, structure determines stellar populations (but with some scatter). 
One implication of this would be that galaxy structure (as traced by effective radius and/or \Sersic\ index) responds more quickly to changes in halo mass than does stellar colour, which would imply that the structural transition from disk-dominated to bulge-dominated precedes the colour transition from blue to red.
By contrast, it would seem that halo mass does not play a primary role in determining the instantaneous star formation rate --- or at least, SSFR is not a good predictor of $M\halo$ --- in this mass range.
By adopting stellar mass as the independent variable or regressor, however, the implicit assumption in the above is that stellar mass can be taken as a proxy for something like `degree of evolution'.

A more theory-minded view would instead cast halo mass play in this primary role. 
In this framing, our stellar mass selection might be viewed as mixing halo populations such that a higher halo mass is offset by a lower stellar-to-halo mass ratio, or vice versa.
The interpretation would then be that, {\em at fixed halo mass:} a more concentrated stellar structure in the present day is associated with relatively less stellar mass; conversely, a more extended and/or diskier stellar distribution is associated with a relatively larger stellar mass.
From this point of view, the implication of our results would be that the processes of star formation and/or stellar assembly have gone slower (through differences in environment, merger history, stellar feedback, or internal dynamics) and/or ended sooner (through differences in environment, merger history, AGN feedback, or internal dynamics), to produce the relatively lower stellar-to-halo mass ratios for generically early types.
\looseness-1

The crucial question thus remains: what is the nature of the astrophysical causal connection(s) underpinning the observed statistical correlations?
From the observers' side, one avenue for further study is the degree to which the dispersion around the SHMR is `random'---in the sense that it is the product of stochastic processes that are not closely correlated with other halo particulars like formation time or large scale environment---or if it instead reflects some form of assembly bias \citep[see, \eg,][]{Wang2013, AlpaslanTinker2020}.
We have no direct means of probing this question with the present data/analysis---instead, what is needed are models that reproduce our results, which can then be interrogated to see how this behaviour comes about in the models. 

In cosmological models of galaxy formation, the mechanism for long-term quenching of star formation in massive galaxies is by heating the gas in the outer halo, and thereby preventing or disrupting further gas accretion onto the galaxies. 
The subgrid prescription for this `maintenance mode' feedback is usually, but not uniquely, associated with kinetic and/or energetic feedback from the central black hole \citep[see, \eg,][]{Dave2016}.
To the extent that AGN feedback scales with black hole mass, and black hole mass scales with bulge mass, and bulge-to-total ratio scales with \Sersic\ index, our results might be used as an indirect test of AGN feedback prescriptions.
For a different perspective, our results would seem at least superficially consistent with, \eg,\ \citet{Snyder2015}, who show that within $10^{12}$ M\sol\ halos from the IllustrisTNG simulations, diskier galaxies tend to have higher stellar masses, and also \citet{Tachella2019}, who argue that present day structure is set during the star-forming phase, where stellar feedback is more important. 
Following this line of argument, our results might be a more sensitive test of stellar feedback prescriptions (but modulo any effects of assembly bias).
In this spirit, we present our observations as targets for modellers to aim to reproduce. \looseness-1

With this first exploratory study, we have demonstrated the feasibility and utility of unstacked lensing profiles to probe variations in halo mass across an ensemble. 
Looking ahead, the obvious next question is whether similar trends exist for lower and higher masses.
This will need more work, and possibly also larger samples to obtain sufficient signal in the lensing measurements.
Taking a broader perspective, this study also shows the value of having galaxy demographic survey data as a foreground screen for wide area lensing surveys.  
In this, we particularly highlight the opportunities that will be afforded by the combination between KiDS and WAVES-Wide \citep{Driver2019}, as well as between Euclid a proposed 4MOST hemisphere survey, in the next few years.

\section*{Acknowledgements}
This research is partially funded by the Australian Government through an Australian Research Council Future Fellowship (FT150100269) awarded to ENT. 
H.\ Hoekstra acknowledges support from Vici grant 639.043.512, financed by the Netherlands Organisation for Scientific Research (NWO). 
AS acknowledges funding from the European Union's Horizon 2020 research and innovation programme under grant agreement No 792916. 
This work is part of the Delta ITP consortium, a program of the Netherlands Organisation for Scientific Research (NWO) that is funded by the Dutch Ministry of Education, Culture and Science (OCW). 
H.\ Hildebrandt is supported by a Heisenberg grant of the Deutsche Forschungsgemeinschaft (Hi 1495/5-1) as well as an ERC Consolidator Grant (No. 770935).
CS acknowledges support from the Agencia Nacional de Investigaci\'on y Desarrollo (ANID) through FONDECYT Iniciaci\'on grant no. 11191125. 

\section*{Data Availability}

The data underlying this article are or soon will be made available through GAMA and KiDS public data releases: see http://gama-survey.org and http://kids.strw.leidenuniv.nl .  The derived results shown in this paper, in the form of MCMC chains, will be shared on reasonable request to the corresponding author. 

\setlength{\bibhang}{2.0em}
\setlength\labelwidth{0.0em}

\end{document}